\documentclass[published]{JHEP}

\usepackage{amssymb}
\usepackage{epsfig}
\usepackage{bbm}
\usepackage{latexsym}

\newcommand{\diag}{\mathop{\rm diag}\nolimits}

\newcommand{\fii}{\varphi_i}
\newcommand{\fj}{\varphi_j}
\def\identity{\mathbbm{1}}

\def \Dbar{D \hskip -7.0pt/\hskip +3.0pt}
\def \dbar{\partial \hskip -5.5pt/\hskip +3.0pt}
\def \lta {\lesssim}

\newcommand{\pt}{\psi_i^T}
\newcommand{\ph}{\theta}
\newcommand{\pk}{k_i \, \psi_i}
\newcommand{\ga}{\bar{\gamma}}
\newcommand{\mpl}{M_{\mathrm P}}
\newcommand{\da}{\dot{a}}
\newcommand{\al}[1]{{\tilde \alpha}_{#1}}

\title{Coupled fields in external background\\ with application to
non-thermal\\ production of gravitinos}

\author{Hans Peter Nilles, Marco Peloso\\ Physikalisches Institut,
Universit{\"{a}}t Bonn\\ Bonn, Germany, Nussallee 12,
D-53115\\ E-mail:
\email{nilles@dirac.physik.uni-bonn.de},\\
\email{peloso@th.physik.uni-bonn.de}}

\author{Lorenzo Sorbo\\ SISSA/ISAS, Trieste, Italy, via Beirut
2-4, I-34013\\ INFN Trieste, Italy, via Valerio 2,
I-34127\\ E-mail: \email{sorbo@sissa.it}}

\abstract{We provide the formalism for the quantization
of systems of coupled bosonic and fermionic fields in a
time dependent classical background. The occupation
numbers of the particle eigenstates can be clearly
defined and computed, through a generalization of the
standard procedure valid for a single field in which
Bogolyubov coefficients are employed. We apply our
formalism to the problem of non-thermal gravitino
production in a two-fields model where supersymmetry is
broken gravitationally in the vacuum. Our explicit
calculations show that this production is strongly
suppressed in the model considered, due to the weak
coupling between the sector which drives inflation and
the one responsible for supersymmetry breakdown.}


\received{March 28, 2001}

\accepted{April 03, 2001}

\JHEP{04(2001)004}

\begin{document}

\section{Introduction}

The analysis of quantized systems in a classical
background can be very useful for the study of various
phenomena that arise in quantum theories, as for example
particle production. The study of matter in external
electromagnetic fields~\cite{itzzub} dates back to the
first years of quantum field
theory~\cite{heieul,schwinger}. For what concerns
gravity~\cite{birdav}, the semiclassical approximation
is often compulsory, due to the lack of a consistent
quantum theory. Despite of this, it turned out very
successful in describing phenomena as particle creation
from black holes~\cite{hawking} or the generation of the
perturbations in the inflationary Universe~\cite{mfb}.

In the last ten years, this semiclassical approach has
been applied to non-thermal particle production after
inflation. In this case, the classical background is
given by the inflaton field, which is coherently
oscillating about the minimum of its potential. The
first analyses of this phenomenon were performed
in~\cite{tb}, but its full relevance was appreciated
only a few years later in the case of production of
scalars~\cite{kls1,stb,kls}. In the work~\cite{kls1}
this non-perturbative production has been called
``preheating'', since it is usually followed by a
ordinary phase of perturbative reheating. It has been
there understood that preheating of bosons is
characterized by a very efficient and explosive
creation, even when single particle decay is
kinematically forbidden. This is due to the coherent
inflaton oscillations, which allow stimulated particle
production into energy bands with very large occupation
numbers.

Less attention was initially paid to non-perturbative
production of fermions. Indeed, the efficiency of this
process seems to be strongly limited by Pauli blocking,
which does not allow for occupation numbers bigger than
one. However, also this production turned out very
relevant, as the first complete calculation~\cite{gprt}
of the inflaton decay into heavy (spin $1/2$) fermions
during preheating showed.\footnote{Among other
interesting studies on production of fermions (not all
of them related to preheating) we
mention~\cite{mamaev,fermioni}.} Indeed, if one only
considers the most natural interactions $\phi \, {\bar
\psi} \, \psi$ and $\phi^2 \, \chi^2$ of the inflaton
$\phi$ to fermions $\psi$ or to bosons $\chi$,
fermionic production occurs in a mass range much broader
than the one for heavy bosons, and this can
``compensate'' the limit imposed by Pauli
blocking~\cite{gprt}.

These results were soon applied~\cite{kklv1,grt1} to
non-thermal gravitino production, since the equations for
the different components of the gravitino field can be
reduced to the one of a spin $1/2$ particle. As had also
been realized in~\cite{mar}, the transverse gravitino
component is always very weakly coupled to the
background and decoupled from the other fermions, so
that the production of its quanta is negligible.
However, the works~\cite{kklv1,grt1} also studied the
production of the longitudinal component, concluding
that it easily exceeds the limits imposed by primordial
nucleosynthesis (the so called ``gravitino
problem''~\cite{grapr}). The analyses
of~\cite{kklv1,grt1} were extended in~\cite{grt2,kklv2}
and followed by several related
works~\cite{granont,allahv}.

Most of these analyses support the conclusions
of~\cite{kklv1,grt1} of a gravitino overproduction at
preheating. However, in all the explicit calculations,
only the case of one chiral superfield with
supersymmetry unbroken in the vacuum was considered.
This last issue is however crucial to understand the
gravitino production: in the super-higgs mechanism (in
the unitary gauge), the gravitino longitudinal component
is provided by the goldstino, which is present only when
supersymmetry is broken. As pointed out in the
works~\cite{grt2,kklv2}, during the cosmological
evolution supersymmetry is broken both by the kinetic
and the potential energies of the scalar fields of the
theory. However, these fields are expected to be settled
in their minima now. If in these minima supersymmetry is
unbroken the gravitino has only the transverse component
and it is massless. The calculations performed in these
schemes show that one fermionic component is produced at
preheating. However, we have remarked that at late times
it cannot be the longitudinal gravitino component. Since
this fermion is the partner of the inflaton field (in
case of only one chiral multiplet the scalar is
necessarily the inflaton), it should be better denoted
as ``inflatino''. Since this field does not have a
gravitational decay rate,\footnote{As also remarked
in~\cite{allahv}, supersymmetry requires the inflatino
decay rate to be comparable with the inflaton one, such
that the inflatino is expected to decay at reheating
(this is actually true provided that supersymmetry
breaking at that time is sufficiently small). In
ref.~\cite{allahv}, explicit calculations are however
performed only with one relevant superfield and unbroken
supersymmetry in the vacuum.} we conclude that
preheating does not contribute to the gravitino problem
in models with supersymmetry unbroken in the vacuum.

One is thus led to consider schemes where both the
issues of inflation at early times and supersymmetry
breaking today are included. From COBE normalization of
the scalar metric perturbation, the relevant scale of
inflation (when only one field is present) is expected
to be about $10^{13}\,\mathrm{GeV}$. Supersymmetry provides
instead a solution to the hierarchy problem if it is
broken at about the TeV scale. Although in principle one
may construct a model where a unique field satisfies
both these requirements, we do not consider this option
as the most natural one. Moving to the two fields case,
the simplest possibility is to consider two well
separate sectors, one of which drives inflation, while
the second is responsible for supersymmetry breaking
today. We have in mind a situation in which no direct
coupling is present between the two fields in the
superpotential, so to have a strong suppression in the
interactions between the inflaton and the field which
provides the longitudinal gravitino component. As we
describe below, with two chiral supermultiplets the
longitudinal gravitino component is coupled to one other
fermionic field (the matter component orthogonal to the
goldstino). For simplicity, in the rest of the
introduction we refer to these two fields simply as to
the gravitino and to the inflatino, although this
interpretation is true only in the vacuum of the theory
and more rigorous definitions will be provided below.
The coupled system is particularly involved, so that
trying to guess its behavior without an explicit
calculation results very difficult. One can guess that
at the preheating era only the inflatino field is
produced. However, there is the potential worry that
much later (on a physical time-scale of the order the
inverse gravitino mass, when supersymmetry is equally
broken by the two sectors of the theory) a fraction of
inflatinos is ``converted'' in gravitinos. As remarked
in~\cite{kklv2}, this worry requires an explicit
calculation in a full supergravity context.

The problem of gravitino production is thus reduced to
the problem of a (quite involved) coupled system in the
external background constituted by the scalar fields of
the theory. The most difficult part of this analysis is
to provide a formalism in which the coupled system is
quantized, with a clear definition of the occupation
numbers for the physical eigenstates. This is a very
interesting problem in itself, which can have several
other applications besides the one we will consider in
the present paper. In the one field case, the procedure
is well established~\cite{mamaev,zeldovich}. One first
quantizes the system and expands the canonical
hamiltonian in the creation and annihilation operators
of the field. The evolution of the background creates a
mixing between the positive and negative energy
solutions of the field equation, which has the
consequence of driving the hamiltonian non diagonal,
even if one takes it to be diagonal at initial time. A
diagonal form is achieved through a (time dependent)
redefinition of the creation/annihilation of the fields.
The two coefficients of this diagonalization are known
as Bogolyubov coefficients and can be easily related to
the occupation number for the quantized field
(consistency of the quantization requires a relation
between these coefficient; this condition is however
preserved by the equations of motion of the system).

In the first part of the present work we generalize this
procedure for systems of multi-fields, both in the
bosonic and in the fermionic case. Although far from
trivial, this generalization can be presented in a
remarkably simple form. By choosing a suitable expansion
of the fields we can repeat each step of the above
analysis substituting the Bogolyubov coefficients with
two matrices $\alpha$ and $\beta$ (the expansion is now
performed in a basis of creation/annihilation operators,
each corresponding to a physical eigenstate of the
system after the diagonalization of the hamiltonian). We
can obtain a system of first order differential
equations for these matrices; the condition on $\alpha$
and $\beta$ for a consistent quantization are also very
simple and are preserved by these equations. Finally,
the expression for the occupation numbers is an easy
generalization of the one valid in the one field
case.\footnote{As in the one field case, a state of
vanishing initial occupation number can be easily
defined, provided the system evolves adiabatically at
initial time.} This first part is divided in two
sections, the first of which is devoted to bosons, while
the second one to fermions. This second section is
further divided in two parts. In the first one we
consider the case in which the fermionic fields are
coupled only through the ``mass matrix'', while in the
second one we consider a more general system which is
necessary for the application to the gravitino case.

The second part of the work is completely devoted to this
application. In section~\ref{sub1} we introduce the quantities
relevant for the calculation, following the notation
of~\cite{kklv2}.  In section~\ref{sub2} we describe the model we
are considering, discussing the evolution of the scalar fields.
In section~\ref{sub3} we write the effective lagrangian for the
longitudinal gravitino component and the matter fermion in the
two chiral supermultiplets case. In the last two subsections we
present our results for the gravitino and the inflatino
production. In section~\ref{sub4} we present analytical results
for the case in which supersymmetry is unbroken in the vacuum,
while numerical results for the case of broken supersymmetry are
shown in section~\ref{sub5}. As we will see, the final gravitino
production is strongly sensitive to the size of the supersymmetry
breaking: our numerical results indicate that this production
goes to zero in the limit of a vanishing final supersymmetry
breaking. This limit is in agreement with the analytical result
of section~\ref{sub4} and with the fact that the longitudinal
gravitino component is actually absent when supersymmetry is
preserved. We thus conclude that in the model considered here,
non-thermal gravitino production is strongly suppressed. This
appears as the consequence of the weak coupling between the two
sectors responsible for inflation and supersymmetry breakdown,
and of the strong hierarchy between the two scales that
characterize them.

A short description of our results for the gravitino and
inflatino production can be found in~\cite{noi}.

\section{System of coupled bosonic fields} \label{secbos}

In this section we consider the coupled system of $N$
bosonic fields $\{ \phi_i \}$ in a FRW background
described by the action
\begin{equation}\label{act1}
S = \frac{1}{2} \, \int d^4 x \, \sqrt{- \, g}  \Big[
\partial_\mu \phi_i \, \partial^\mu \phi_i - m_{i\,j}^2
\phi_i \, \phi_j + \xi \, R \, \phi_i \phi_i \Big]\,.
\end{equation}
We use conformal time $\eta$, such that the metric and the Ricci
scalar are $g_{\mu \, \nu} = a^2 ( \eta ) \diag (1,\, -1,\, -1,\,
-1 )$ and $R = - \, 6 \, \ddot{a} / a^3$, where $a$ is the scale
factor of the Universe and dot denotes derivative with respect to
$\eta$ (summation over repeated indices is understood). The last
term describes a possible non-minimal coupling ($\xi \neq 0$) of
the scalar fields to gravity.

The (symmetric) mass matrix $m_{i\,j}^2$ is assumed to
be a function of some external (background) fields. The
only assumption that we do on these external fields is
that they are constant (or better, adiabatically
evolving) at the very beginning\footnote{We require an
initial stage of adiabatic evolution to consistently
define vanishing occupation numbers for the bosons at
initial time.} and at the very end of the evolution of
the system. In these regimes, the matrix $m_{i\,j}^2$
becomes also constant and the fields which diagonalize
it become free fields, whose masses are precisely given
by the eigenvalues of $m_{i\,j}^2$. However, during the
evolution the different entries of $m_{i\,j}^2$ are
allowed to vary, and the (time dependent) eigenvalues of
$m_{ij}^2$ are interacting fields whose masses change in
time. These masses can change non adiabatically and this
may in general lead to particle production. The aim of
this section is to give a precise definition of the
occupation number and to provide the formalism to
calculate it.

It is most convenient to consider the ``comoving''
fields $\fii \equiv a\, \phi_i$. For these fields, the
above action~(\ref{act1}) rewrites\footnote{We do not
necessarily need a cosmological motivation for the
analysis that we perform in the rest of this section.
The action~(\ref{act2}) could indeed also arise in flat
space, with a non-diagonal $\Omega_{ij}^2$ coming from
some general interactions between the bosons $\fii$ and
some other fields which have been integrated out.}
\begin{eqnarray}
S &=& \frac{1}{2} \int d^4 x \left[ \dot{\fii} \,
\dot{\fii} - \fii \, \Omega_{ij}^2 \, \fj \right], \nonumber\\
\Omega_{ij}^2 &\equiv& a^2 \, m_{ij}^2 + \left( - \Delta +
\frac{\ddot{a}}{a} \left( 6 \, \xi - 1 \right) \right)
\delta_{ij}\,,\label{act2}
\end{eqnarray}
where $\Delta$ is the laplacian operator. We can also
write the hamiltonian of the system, which, in terms of
the fields $\fii$ and their conjugate momenta
\begin{equation}
\Pi_i \equiv \frac{\partial \, \mathcal{L}}{\partial \dot{\fii}}
= \dot{\fii}\,,\label{conjbos}
\end{equation}
reads
\begin{equation}
H \equiv \int d^3 \mathbf{x} \, \mathcal{H} =
\frac{1}{2} \int d^3 \mathbf{x} \left( \Pi_i \, \Pi_i
+ \fii \, \Omega_{ij}^2 \fj \right).\label{hambos}
\end{equation}

The frequency matrix $\Omega_{ij}^2$ which enters in the
above expressions is also in general time dependent and
non diagonal. At any given time, it can be diagonalized
with an orthogonal matrix $C$
\begin{equation}
C^T \left( \eta \right) \Omega^2 \left( \eta \right) C
\left( \eta \right) = \omega^2 \left( \eta \right) \qquad
\mbox{diagonal}\,.\label{omdbos}
\end{equation}
We denote by ${\hat \varphi} \equiv C^T \, \varphi$ the
bosonic fields in the basis in which the frequency
matrix is diagonal. We also denote by $\omega_i^2$ the
$i$-th entry of the diagonal matrix $\omega^2$. The
set of $\omega_i$ represents the energy densities of the
(time dependent) physical eigenstates of the system
${\hat \fii}$.

We now show that the occupation numbers of these fields
can be defined and computed generalizing the usual
techniques based on Bogolyubov coefficients valid in the
one field case. The first step to do this is to consider
a basis for annihilation/creation operators $\{a_i \}$
and $\{a_i^\dagger \}$ and to perform the decompositions
\begin{eqnarray}
\fii &=& C_{ij} \int \frac{d^3 k}{\left( 2 \, \pi \right)^{3/2}}
\left[ \mathrm{e}^{i\, \mathbf{k} \cdot \mathbf{x}} \, h_{jk}
\left( \eta \right) a_k \left( k \right) +
\mathrm{e}^{- i\, \mathbf{k} \cdot \mathbf{x}} \, h_{jk}^*
\left(\eta \right) a_k^\dagger \left( k \right) \right],
\nonumber\\
\Pi_i &=& C_{ij} \int \frac{d^3 k}{\left( 2 \, \pi \right)^{3/2}}
\left[ \mathrm{e}^{i\, \mathbf{k} \cdot \mathbf{x}} \,
{\tilde h_{jk}}\left( \eta \right) a_k \left( k \right)
+\mathrm{e}^{-i\,\mathbf{k} \cdot \mathbf{x}} \,
{\tilde h_{jk}^*} \left( \eta \right)a_k^\dagger \left( k \right)
\right].\label{canbos}
\end{eqnarray}
The reason why we explicitly factorized the matrix $C$
in these decompositions will be clear soon. Due to the
fact that the fields are coupled together, the matrices
$h$ and ${\tilde h}$ are generically expected to be non
diagonal.

To quantize the system, we impose
\begin{equation}
\left[ \fii \left(x \right),\, \Pi_j \left( y \right) \right]
= i \, \delta^3 \left( \mathbf{x} - \mathbf{y} \right)
\delta_{ij}\label{canon}
\end{equation}
for the conjugate fields, and
\begin{equation}
\left[ a_i \left( \mathbf{k} \right), a_j^\dagger
\left(\mathbf{p} \right) \right] = \delta^3 \left(\mathbf{k}
- \mathbf{p} \right) \delta_{ij}\label{etcr}
\end{equation}
for the annihilation/creation operators. We can satisfy
both these relations requiring
\begin{equation}
\left[h \, {\tilde h^\dagger} - h^* \, {\tilde h}^T \right]_{ij}
= i \, \delta_{ij}\,,\label{relh}
\end{equation}
as it can be easily checked from the
decomposition~(\ref{canbos}).

From the action~(\ref{act2}), one deduces the second
order equations of motion
\begin{equation}
\ddot{\fii} + \Omega_{ij}^2 \, \fj = 0\,. \label{eombos}
\end{equation}
However, one can achieve a system of only first order equations
by setting some additional relations between the conjugate fields
$\fii$ and $\Pi_i$. We want these relations to generalize the one
which is usually taken in the one field case, see i.e.\
\cite{zeldovich}. Also we want them to allow a rewriting of the
hamiltonian~(\ref{hambos}) in a simple and readable form. The
sets of fields where this generalization is most evident is given
by $\{ {\hat \fii}, {\hat \Pi_i} \equiv ( C^T \, \Pi )_i \}$.
These fields are decomposed as in eqs.~(\ref{canbos}), only
without the $C$ matrix before the integrals. In terms of these
fields, the hamiltonian~(\ref{hambos}) rewrites
\begin{equation}
H = \int d^3 \mathbf{x} \; \frac{1}{2} \left( {\hat
\Pi_i}\, {\hat \Pi_i} + \omega_i^2 \, {\hat \fii}
\, {\hat \fii} \right),\label{hambos2}
\end{equation}
since the frequency $\omega$ is diagonal. One is thus led to
impose the conditions\footnote{Equations~(\ref{posiz}) are written
in matrix notation. In general, for any function $f ( \omega_i )$
and any matrix $M$, we use the notation
\begin{equation}
\left(f \left( \omega \right) M \right)_{ij} \equiv f
\left( \omega_i \right) M_{ij}\,,\qquad \left(M \, f
\left( \omega \right) \right)_{ij} \equiv M_{ij} \,
f \left( \omega_j \right).
\label{nota}
\end{equation}.}
\begin{eqnarray}
h &=& \frac{\mathrm{e}^{-i \, \int^\eta \omega \, d
\eta'}}{\sqrt{2 \, \omega}} \, A +  \frac{\mathrm{ e}^{i
\, \int^\eta \omega \, d \eta'}}{\sqrt{2 \, \omega}} \,
B\,, \nonumber\\ {\tilde{h}} &=& \frac{- \, i \,
\omega \, \mathrm{e}^{-i \, \int^\eta \omega \, d
\eta'}}{\sqrt{2 \, \omega}} \, A +  \frac{i \omega \,
\mathrm{e}^{i \, \int^\eta \omega \, d \eta'}}{\sqrt{2
\, \omega}} \, B \label{posiz}
\end{eqnarray}
which are indeed a natural generalization of the one
which is usually taken in the one field
case~\cite{zeldovich}. For one field, $A$ and $B$ are
numbers, known as Bogolyubov coefficients. In our case
they are $N \times N$ matrices. The analysis of the
system is in our case simplified if we consider, rather
then the matrices $A$ and $B$, the combinations
\begin{eqnarray}
\alpha &\equiv& \mathrm{e}^{-i\,\int^\eta \omega \, d
\eta'} \, A\,, \nonumber\\
\beta &\equiv& \mathrm{e}^{i\,\int^\eta \omega \,d \eta'} \, B\,.
\end{eqnarray}

The above condition~(\ref{relh}) is satisfied if the
matrices $\alpha$ and $\beta$ obey the relations
\begin{eqnarray}
\alpha\,\alpha^\dagger-\beta^* \,\beta^T &=& 1\,, \nonumber\\
\alpha\,\beta^\dagger-\beta^*\,\alpha^T &=& 0\,.\label{bogbosc}
\end{eqnarray}
These relations can be imposed at the initial time, and
are preserved by the evolution, as we shortly discuss.
In the one field case, they reduce to the usual
condition $\vert \alpha \vert^2 - \vert \beta \vert^2 =
\vert A\vert^2 - \vert B \vert^2 = 1$.

As we have said, the evolution of the system can be
described by two sets of first order differential
equations. The first set is obtained by inserting
eqs.~(\ref{canbos}) into the definition of the conjugate
momenta, eq.~(\ref{conjbos})
\begin{equation}
{\dot h} = {\tilde h} - \Gamma \, h\,,\label{eqb1}
\end{equation}
where we have defined the matrix
\begin{equation}
\Gamma = C^T \, \dot{C}\,,\qquad \Gamma^T = - \Gamma \,.
\end{equation}
The second set of equations is obtained by rewriting
eqs.~(\ref{eombos}) in terms of $\fii$ and~$\Pi_i$
\begin{equation}
\dot{\tilde h} = - \Gamma \, {\tilde h} - \omega^2 \, h \,.
\end{equation}

We can now use relations~(\ref{posiz}) and decouple the
terms proportional to $\dot{\alpha}$ and $\dot{\beta}$,
so to arrive to the final result
\begin{eqnarray}
\dot{\alpha} &=& - i \, \omega \, \alpha +
\frac{\dot{\omega}}{2 \, \omega} \, \beta - I \,
\alpha - J \, \beta \,, \nonumber\\
\dot{\beta} &=& \frac{\dot{\omega}}{2 \, \omega} \,
\alpha +  i \, \omega \, \beta - J \, \alpha - I \,
\beta \,,\label{eqfinbos}
\end{eqnarray}
where we have defined the matrices
\begin{eqnarray}
I &=& \frac{1}{2}  \left( \sqrt{\omega} \, \Gamma \,
\frac{1}{\sqrt{\omega}} + \frac{1}{\sqrt{\omega}} \,
\Gamma \, \sqrt{\omega} \right),\qquad
I^T = - I \,, \nonumber\\
J &=& \frac{1}{2} \left(\sqrt{\omega} \, \Gamma \,
\frac{1}{\sqrt{\omega}} -\frac{1}{\sqrt{\omega}} \, \Gamma \,
\sqrt{\omega}\right),\qquad J^T = J \,.
\end{eqnarray}

In the one field case, $I=J=\Gamma = 0$, and the above
system reduces to the equations for the two Bogolyubov
coefficients
\begin{equation}
\dot{A} = \frac{\dot{\omega}}{2 \, \omega} \,
\mathrm{e}^{2 \, i \, \int^\eta\omega \, d \eta'} B\,,\qquad
\dot{B} = \frac{\dot{\omega}}{2 \,\omega} \, \mathrm{e}^{-2 \,
i \, \int^\eta \omega \, d \eta'} A \,,\label{eqabbos}
\end{equation}
already discussed in the previous literature (see i.e.\
\cite{kls}). In the one field case the only source of
nonadiabaticity is related to a rapid change of the only
frequency $\omega ( \eta )$, so that the system is said to evolve
adiabatically whenever the condition $\dot{\omega} \ll
\omega^2$ is fulfilled. In the present case, there are more
sources of nonadiabaticity, related to the fact that now the
frequency $\Omega_{ij}$ is a $N \times N$ matrix. This is
associated with the presence of non-vanishing matrices $I$ and
$J$ in the equations of motion for the matrices $\alpha$ and
$\beta$.

It is a straightforward exercise to show that the above
equations~(\ref{eqfinbos}) preserve the normalization
conditions~(\ref{bogbosc}), due to the properties $I^T =
-  I$ and $J^T = J$.

In the one field case, the number of particles is given
by the modulus square of the second Bogolyubov
coefficient, $\vert B \vert^2$. We now show that also in
the multi-field case it is generally related to the
matrix $\beta$. \pagebreak[3] To see this, we decompose also the
energy density operator $\mathcal{H}$ (see
eq.~(\ref{hambos})) in the basis of annihilation and
creation operators
\begin{equation}
\mathcal{H} = \left(a_i^\dagger,\,  a_j \right)
\pmatrix{\mathcal{E}_{il} & \mathcal{F}_{jl}^\dagger \cr
\mathcal{F}_{im} & \mathcal{E}_{jm}^T}
\pmatrix{a_l \cr\ a_m^\dagger}.\label{hdecbos}
\end{equation}
From eqs.~(\ref{canbos}), one sees that the $N \times N$
matrices $\mathcal{E}$ and $\mathcal{F}$ which enter in
this decomposition are given by
\begin{eqnarray}
\mathcal{E} &=& \frac{1}{2} \left( {\tilde h}^\dagger \,
{\tilde h} + h^\dagger \, \omega^2 \, h \right),\nonumber\\
\mathcal{F} &=& \frac{1}{2} \left( {\tilde h}^T \, {\tilde h}
+ h^T \, \omega^2 \, h \right).\label{efbos}
\end{eqnarray}

We can now generalize the procedure adopted in the one
field case. The matrix that appears in
eq.~(\ref{hdecbos}) can be put in diagonal form in a
basis of new (time dependent) annihilation/creation
operators. Only when the hamiltonian is diagonal, each
pair of (redefined) operators can be associated to a
physical particle, and used to compute the corresponding
occupation number. The explicit computation gives
\begin{eqnarray}
\mathcal{E} &=& \frac{1}{2} \left(\alpha^\dagger \, \omega \,
\alpha + \beta^\dagger \, \omega \, \beta \right), \nonumber\\
\mathcal{F} &=& \frac{1}{2} \left(\alpha^T \, \omega \, \beta
+ \beta^T \, \omega \, \alpha \right),
\end{eqnarray}
so that we found that expression~(\ref{hdecbos})
evaluates to
\begin{equation}
\mathcal{H} = \frac{1}{2} \left(a^\dagger,\, a \right)
\pmatrix{\alpha^\dagger &\beta^\dagger \cr \beta^T & \alpha^T}
\pmatrix{\omega & 0 \cr 0 & \omega}
\pmatrix{\alpha & \beta^* \cr \beta & \alpha^*}
\pmatrix{a \cr a^\dagger}.\label{hadbos}
\end{equation}
In terms of the redefined annihilation/creation
operators\footnote{The relation~(\ref{physopbos}) is
inverted through the matrix
\begin{equation}
\pmatrix{\alpha^\dagger & - \beta^\dagger \cr - \beta^T &
\alpha^T},
\end{equation}
as can be easily checked from
conditions~(\ref{bogbosc}). We thus see that also the
relations
\begin{equation}
\alpha^\dagger \, \alpha - \beta^\dagger \, \beta = 1 \,,\qquad
\alpha^\dagger \, \beta^* - \beta^\dagger \, \alpha^* = 0
\end{equation}
hold for the whole evolution.}
\begin{equation}
\pmatrix{{\hat a} \cr {\hat a}^\dagger} \equiv
\pmatrix{\alpha & \beta^* \cr \beta & \alpha^*}
\pmatrix{a \cr a^\dagger}\label{physopbos}
\end{equation}
the hamiltonian is thus diagonal (remember that in
eq.~(\ref{omdbos}) $\omega$ was defined to be diagonal),
and, after normal ordering, it simply reads
\begin{equation}
H = \int d^3 \, \mathbf{k} \: \omega_i \, {\hat
a}_i^\dagger \, {\hat a}_i \,.
\end{equation}

We choose at initial time $\alpha ( \eta_0 ) =
\identity,\ \beta ( \eta_0 ) = 0$, so that
conditions~(\ref{bogbosc}) are fulfilled. We also choose the
initial state of the theory to be annihilated by the operators
$a_i$. At any generic time, the occupation number of the $i$-th
bosonic eigenstate is given by (notice that in this expression we
do not sum over $i$)
\begin{equation}
N_i \left( \eta \right) = \langle {\hat a}_i^\dagger \,
{\hat a}_i \rangle = \left( \beta^* \beta^T \right)_{ii}.
\label{numbos}
\end{equation}
In the one field case the above relation reduces to the
usual $N = \vert B \vert^2$. We see that our choices
correspond to an initial vanishing occupation number for
all the bosonic~fields.

\section{System of coupled fermionic fields}

We now consider a system of coupled fermions. We divide
this analysis in two subsections. The first of them
extends to the fermionic case the results obtained for
bosons in the previous section. Because of the repeated
analogies, the discussion is here shorter than the above
one, where more details can be found. In the second
subsection we study a more general system of equations,
which can be also relevant when the background is not
constant. In particular, these can be important for
cosmology, where the expansion of the Universe provides
a preferred direction in time. In the next section we
will indeed discuss non-perturbative production of
gravitinos as an application of this analysis.

\subsection{The simpler case}

Let us consider the coupled system of $N$ fermionic
fields $\{ \psi_i \}$ in a FRW background described by
the action
\begin{equation}
S = \int d^4 x \, \sqrt{- g} \, {\bar \psi}_i \left[
i \, \delta_{ij} \, \left( {\tilde \gamma^\mu} \,
\partial_\mu + \frac{3}{2} \, \frac{\dot{a}}{a} \,
{\tilde \gamma^0} \right) - M_{ij} \right] \psi_j\,.\label{act1fer}
\end{equation}
The gamma matrices ${\tilde \gamma^\mu}$ in FRW geometry
are related to those ($\gamma^\mu$) in flat space by
${\tilde \gamma^\mu} = a^{-1} \,\gamma^\mu$, where $a
\left( \eta \right)$ is the scale factor of the
Universe. As before, conformal time $\eta$ is used, and the
matrix $M_{ij}$ is considered to be a function of some external
background fields. The requirement that the action is hermitean
forces $M$ to be hermitean as well. For simplicity we will also
take it to be real. We also require $M_{ij}$ to be constant
(better, adiabatically evolving) at very early and late times,
but we do not make any other assumption on its evolution.

After the redefinitions $X_i \equiv \psi_i \, a^{3/2}$, $m
\equiv a \, M$, the action~(\ref{act1fer}) reads
\begin{equation}
S = \int d^4 x \, {\bar X}_i \left[ i \, \delta_{ij} \,
\gamma^\mu \,  \partial_\mu - m_{ij} \right] X_j\,,
\label{act2fer}
\end{equation}
leading to the equations of motion (in matrix notation)
\begin{equation}
\left( i \, \gamma^\mu \, \partial_\mu - m \right) X = 0\,.
\label{eomferm}
\end{equation}
The canonical hamiltonian is instead
\begin{equation}
H \equiv \int d^3 \mathbf{x} \; \mathcal{H} = \int d^3
\mathbf{x} \, {\bar X} \left[- i \, \gamma^i \,
\partial_i + m \right] X\,.\label{hamferm}
\end{equation}

In analogy with the bosonic case, we expand the
fermionic eigenstates into a basis of
creation/annihilation operators
\begin{equation}
X_i \left( x \right) = C_{ij} \int \frac{d^3
\mathbf{x}}{\left( 2 \pi \right)^{3/2}} e^{i \mathbf{k
\cdot x}} \Big[ U_r^{jk} \left( k,\, \eta \right)
a_r^k \left( k \right) + V_r^{jk} \left( k,\, \eta
\right) b_r^{+ k} \left( - k \right)
\Big]\,,\label{exp1ferm}
\end{equation}
where the matrix $C$ is employed into the
diagonalization of the mass matrix $m$
\begin{equation}
\mu \equiv C^T \, m \, C\,,\qquad C \ \mbox{orthogonal}\,.
\end{equation}

We also define the matrix (dot denotes derivative with
respect to $\eta$)
\begin{equation}
\Gamma \equiv C^T \, \dot{C}\,,\qquad \Gamma^T = - \Gamma \,,
\end{equation}
and the ``generalized spinors''
\begin{equation}
U_r^{ij} \equiv \left[ \frac{U_+^{ij}}{\sqrt{2}} \, \psi_r,
r\,\frac{U_-^{ij}}{\sqrt{2}} \, \psi_r \right]^T,\qquad
V_r^{ij} \equiv \left[ \frac{V_+^{ij}}{\sqrt{2}} \, \psi_r,
r\,\frac{V_-^{ij}}{\sqrt{2}} \, \psi_r \right]^T \label{spin}
\end{equation}
with $\psi_{+} =$ {\scriptsize{$\pmatrix{1 \cr 0}$}} and $\psi_{-}
=$ {\scriptsize{$\pmatrix{0 \cr 1}$}} eigenvectors of the helicity
operator $\mathbf{\sigma}\cdot \mathbf{v} / \vert \mathbf{v}
\vert$.

Let us consider a set of fields $X_i$ which satisfy the
above equations~(\ref{eomferm}). Due to the fact that
the matrix $m_{ij}$ is real and symmetric, then also the
fields $X_i^C \equiv {\tilde C} \, {\bar X_i}^T$
(where ${\tilde C}$ is the charge conjugation
matrix\footnote{In our computations, we take
\begin{equation}
\gamma^0 = \pmatrix{\identity & 0 \cr 0 & - \identity},
\qquad \gamma^i = \pmatrix{0 & \sigma_i \cr - \sigma_i & 0},
\qquad {\tilde C} = i \, \gamma^0 \, \gamma^2 =
\pmatrix{0 & i \, \sigma_2 \cr i \, \sigma_2 & 0},
\end{equation}
where $\sigma$ are the Pauli matrices.}) are solutions
of~(\ref{eomferm}).\footnote{This may allow one to consistently
define the Majorana condition $X_i^C \equiv X_i$.} As a
consequence, one can impose the relation $U_r ( k ) = {\tilde C}
\, {\bar V}_r^T \,
( - k )$, or, using eqs.~(\ref{spin}),
\begin{equation}
V_+ = - U_-^* \,,\qquad V_- = U_+^* \,.\label{majorana}
\end{equation}
Doing so, we have only to deal with the $U_\pm$
matrices. Taking the momentum $k$ along the third axis,
their equations of motion read
\begin{equation}
\dot{U_\pm} = - i \, k \, U_\mp \mp i \, \mu \, U_\pm -
\Gamma \, U_\pm \,.\label{eomu}
\end{equation}

The quantization of the system requires
\begin{eqnarray}
\left\{ X_i \left( \eta,\, \mathbf{x} \right), X_j^\dagger
\left( \eta ,\, \mathbf{y} \right) \right\} &=& \delta^{(3)}
\left( \mathbf{x} - \mathbf{y} \right) \delta_{ij} \,, \nonumber\\
\left\{ a_{r\,i} \left( \mathbf{k} \right),\, a_{s\,j}^\dagger
\left( \mathbf{p} \right) \right\} &=& \delta^{(3)}
\left( \mathbf{k} - \mathbf{p} \right) \delta_{rs} \,
\delta_{ij} \,, \nonumber\\
\left\{ b_{r\,i} \left( \mathbf{k} \right),\, b_{s\,j}^\dagger
\left( \mathbf{p} \right) \right\} &=& \delta^{(3)}
\left(\mathbf{k} - \mathbf{p} \right) \delta_{rs} \,
\delta_{ij} \,.\label{quantferm}
\end{eqnarray}

We can simultaneously satisfy these conditions by
setting\footnote{Notice that all this analysis
generalizes the one made in the one field case. For the
latter, we follow~\cite{mamaev}. A detailed exposition
with a notation closer to the present one is found
in~\cite{ps}.}
\begin{eqnarray}
U_+U_+^\dagger+U_-^* \, U_-^T &=& 2 \, \identity \,,\nonumber \\
U_+ \,U_-^\dagger &=& U_-^* \, U_+^T \,.\label{cond1ferm}
\end{eqnarray}
These conditions can be imposed at initial time, and are
preserved by the evolution of the system (as it is
easily checked from eqs.~(\ref{eomu})).

We define the diagonal matrix\footnote{This definition
is meaningful, since both $\omega$ and $\mu$ are
diagonal matrices. More simply, it can be understood as
a relation between their eigenvalues. See also the
footnote with eq.~(\ref{nota}) for some clarification
about our notation.}
\begin{equation}
\omega = \sqrt{k^2 + \mu^2} \,,
\end{equation}
and we further expand
\begin{eqnarray}
U_+ &\equiv& \left( 1 + \frac{\mu}{\omega} \right)^{1/2}
\mathrm{e}^{- i \int^\eta \omega \, d \eta'} \, A -
\left( 1 - \frac{\mu}{\omega} \right)^{1/2}
\mathrm{e}^{i \int^\eta \omega \, d \eta'} \, B \nonumber\\
&\equiv& \left( 1 + \frac{\mu}{\omega} \right)^{1/2} \alpha
- \left( 1 - \frac{\mu}{\omega} \right)^{1/2} \beta\,,
\nonumber\\
U_- &\equiv& \left(1 - \frac{\mu}{\omega} \right)^{1/2}
\mathrm{e}^{- i \int^\eta \omega \, d \eta'} \, A +
\left( 1 + \frac{\mu}{\omega} \right)^{1/2}
\mathrm{e}^{i \int^\eta \omega \, d \eta'} \, B\nonumber\\
&\equiv& \left( 1 - \frac{\mu}{\omega}
\right)^{1/2} \alpha + \left( 1 + \frac{\mu}{\omega}
\right)^{1/2} \beta \,,\label{decoferm}
\end{eqnarray}
so that the above conditions~(\ref{cond1ferm}) are satisfied if
the matrices $\alpha$ and $\beta$ obey the relations
\begin{eqnarray}
\alpha \, \alpha^\dagger + \beta^* \, \beta^T &=& \identity \,,
\nonumber\\
\alpha \, \beta^\dagger - \beta^* \, \alpha^T &=& 0 \,.
\label{cond2ferm}
\end{eqnarray}
In the one field case, $A$ and $B$ are numbers, known as
Bogolyubov coefficients. In our case they are $N \times
N$ matrices. The matrices $\alpha$ and $\beta$ are
introduced since their equations of motion assume a
simpler form then the corresponding ones for $A$ and
$B$. In the one field case, the above
relations~(\ref{cond2ferm}) reduce to the usual
condition $\vert A \vert^2 + \vert B \vert^2 = 1$.

For fermions, the evolution equations for the matrices
$\alpha$ and $\beta$ can be obtained in a more
straightforward way with respect to the bosonic case.
This is because eqs.~(\ref{eomu}) are already two sets
of first order differential equations. Using the above
decomposition~(\ref{decoferm}), after some algebra we
arrive to the final expressions\footnote{In case of only
one superfield, these equations simplify to
\begin{equation}
\dot{A} = - \frac{{\dot \mu} \, k}{2 \, \omega^2} \,
\mathrm{e}^{2 \, i \, \int^\eta \omega d \eta'}  B \,,\qquad
\dot{B} = \frac{{\dot \mu} \, k}{2 \, \omega^2} \,
\mathrm{e}^{- 2 \, i \, \int^\eta \omega d \eta'}  A \,.
\label{eq1ferm}
\end{equation}}
\begin{eqnarray}
\dot{\alpha} &=& \left[ - i  \, \omega - I \right] \alpha
+ \left[ - \frac{\dot{\mu} \, k}{2 \, \omega^2} + J \right]
\beta \,, \nonumber\\
\dot{\beta} &=& \left[ \frac{\dot{\mu} \, k}{2 \, \omega^2} -
J \right] \alpha + \left[ i \, \omega - I \right] \beta \,,
\label{eomabferm}
\end{eqnarray}
where we have defined the matrices
\begin{eqnarray}
2  I &\equiv& \left(1 + \frac{\mu}{\omega}\right)^{1/2}
\Gamma  \left( 1 + \frac{\mu}{\omega} \right)^{1/2}
+ \left( 1 - \frac{\mu}{\omega} \right)^{1/2}  \Gamma
\left(1 - \frac{\mu}{\omega} \right)^{1/2},\qquad
I^T = -  I \,, \nonumber\\
2  J &\equiv& \left(1 + \frac{\mu}{\omega}\right)^{1/2}
\Gamma  \left(1 - \frac{\mu}{\omega} \right)^{1/2} -
\left(1 - \frac{\mu}{\omega} \right)^{1/2}  \Gamma
\left(1 + \frac{\mu}{\omega} \right)^{1/2},\qquad
J^T = J \,.\quad
\end{eqnarray}

One can easily verify that these equations preserve the
above conditions~(\ref{cond2ferm}).

To properly define and compute the occupation number for
the fermionic eigenstates, as before we expand the
energy density operator $\mathcal{H}$
(eq.~(\ref{hamferm})) in the basis of annihilation and
creation operators
\begin{equation}
\mathcal{H} = \left(a_i^\dagger \,,\,  b_j \right)
\pmatrix{\mathcal{E}_{il} & \mathcal{F}_{jl}^\dagger \cr
\mathcal{F}_{im} & - \mathcal{E}_{jm}^T}
\pmatrix{a_l \cr b_m^\dagger}.\label{hdecferm}
\end{equation}
Using eqs.~(\ref{exp1ferm}) and~(\ref{spin}), we find
\begin{eqnarray}
\mathcal{E} \left( \eta \right) &\equiv& \frac{1}{2}
\Big[ U_+^+ \, \mu \, U_+ - U_-^+ \, \mu \, U_- + U_+^+ \,
k \, U_- + U_+^+ \, k \, U_+ \Big]\,, \nonumber\\
\mathcal{F} \left(\eta \right) &\equiv& \frac{1}{2}
\Big[- U_+^T \, \mu \, U_- - U_-^T \, \mu \, U_+ + U_+^T \,
k \, U_+ - U_-^T \, k \, U_- \Big]\,,
\end{eqnarray}
while eqs.~(\ref{decoferm}) lead to
\begin{eqnarray}
\mathcal{E} &=& \alpha^\dagger \, \omega \, \alpha -
\beta^\dagger \, \omega \, \beta\,, \nonumber\\
\mathcal{F} &=& - \alpha^T \, \omega \, \beta - \beta^T \,
\omega \, \alpha \,.
\end{eqnarray}

We have thus
\begin{equation}
\mathcal{H} = \left( a^\dagger,\, b \right)
\pmatrix{\alpha^\dagger & \beta^\dagger \cr -  \beta^T & \alpha^T}
\pmatrix{\omega & 0 \cr 0 & - \omega}
\pmatrix{\alpha & - \beta^* \cr \beta & \alpha^*}
\pmatrix{a \cr b^\dagger}.\label{hadferm}
\end{equation}

In terms of the redefined annihilation/creation
operators\footnote{The matrix in eq.~(\ref{physopferm})
is unitary, so its inverse one is precisely given by
\begin{equation}
\pmatrix{\alpha^\dagger & \beta^\dagger \cr - \beta^T & \alpha^T},
\end{equation}
as can be easily checked from
conditions~(\ref{cond2ferm}). We thus see that also the
relations
\begin{equation}
\alpha^\dagger \, \alpha + \beta^\dagger \, \beta = \identity \,,
\qquad \alpha^\dagger \, \beta^* - \beta^\dagger \, \alpha^* = 0
\end{equation}
hold for the whole evolution.}
\begin{equation}
\pmatrix{{\hat a} \cr {\hat b}^\dagger} \equiv
\pmatrix{\alpha & -  \beta^* \cr \beta & \alpha^*}
\pmatrix{a \cr  b^\dagger}\label{physopferm}
\end{equation}
the hamiltonian is thus diagonal, and, after normal
ordering, it simply reads
\begin{equation}
H = \int d^3 \, \mathbf{k} \: \omega_i \left( {\hat
a}_i^\dagger \, {\hat a}_i + {\hat b}_i^\dagger \,
{\hat b}_i \right).
\end{equation}

We choose at initial time $\alpha ( \eta_0 ) = 1,\beta(\eta_0 ) =
0$, so that conditions~(\ref{cond2ferm}) are fulfilled. We also
choose the vacuum state of the theory to be annihilated by the
initial annihilation operators $a_i$ and $b_i$. At any given
time, the occupation number of the $i$-th fermionic eigenstate is
given by (notice that in this expression we do not sum over $i$)
\begin{equation}
N_i \left( \eta \right) = \langle {\hat a}_i^\dagger \,
{\hat a}_i \rangle = \langle {\hat b}_i^\dagger \, {\hat
b}_i \rangle = \left( \beta^* \beta^T \right)_{ii}.\label{numferm}
\end{equation}

In the one field case the above relation reduces to the
usual $N = \vert B\vert^2$. We see that our choices
correspond to an initial vanishing occupation number for
all the fermionic fields. Notice that particles and
antiparticles have the same energy and are produced with
the same amount, due to the reality conditions that we
have imposed on the system. Finally, we observe that the
first of conditions~(\ref{cond2ferm}) guarantees that
Pauli blocking is always satisfied.

\subsection{A more general case} \label{subgen}

We now consider a more general action for the coupled
system of $N$ fermionic fields. For future convenience,
here we switch to the signature $-, +, +,+$ for the
Minkowski metric, and we then work with the gamma
matrices
\begin{equation}
\ga^0 = \pmatrix{- i \, \identity & 0 \cr 0 & i \, \identity},
\qquad \ga^i = \pmatrix{0 & -i\, \sigma_i \cr i \, \sigma_i & 0}
\label{gammas}
\end{equation}
in flat space.

By a suitable conformal rescaling of the fermionic
fields and of their masses as we did before
eq.~(\ref{act2fer}), we can again remove the scale
factor of the Universe from the kinetic term for the
fermions. \pagebreak[3] However, we are now interested in a more
generic system, so that we consider, instead
of~(\ref{act2fer}), the action
\begin{equation}
S = \int d^4 x \, {\bar X}_m \left[ \ga^0 \partial_0 +
\ga^i \, N \, \partial_i + M \right]_{mn} X_n
\,,\label{actgenfer}
\end{equation}
where $N$ and $M$ are two $N \times N$ matrices of the
form
\begin{equation}
N \equiv N_1 + \ga^0 \, N_2 \,,\qquad M \equiv M_1 +
\ga^0 \, M_2 \,.
\end{equation}

The matrices $N$ and $M$ are assumed to be functions of
some external fields. We consider a situation in which
these fields evolve in time. This time dependence
justifies the general form for the action that we want
to discuss. As we will see in the next section, this
analysis can have relevance for cosmology, where the
expansion of the Universe provides a natural direction
for time. However, the system~(\ref{actgenfer}) could
also arise in flat space from some general interactions
between the fermions $X_i$ and other fields which have
been integrated out. As in the previous analyses, our
main goal is to discuss the definition of the occupation
number of the physical eigenstates of the system, and to
provide the formalism to calculate it.

We list here our assumptions on the matrices $M$ and
$N$. First, we require them to change adiabatically at
initial times, so to consistently define the initial
occupation numbers. Then, we assume $M_1 \rightarrow
\mbox{constant}, M_2 \rightarrow 0, N
\rightarrow \identity$ at late times, so to recover a
system of ``standard'' decoupled particles at the end (indeed,
one can choose the basis of fields $X_i$ such that the matrix $M$
is diagonal at the end). The requirement of an hermitean action
translates into the conditions
\begin{equation}
N_i^\dagger = N_i \,,\qquad M_1 = M_1^\dagger \,,\qquad
M_2 = - M_2^\dagger \,.
\end{equation}
For simplicity, we will only consider real matrices, so
that $N_1$, $N_2$ and $M_1$ are required to be
symmetric, while $M_2$ antisymmetric. Finally, we impose
an additional condition, which is that the kinetic term
for the fermions ``squares'' to the D'Alambertian
operator $\Box$. If we take the equations of motion
following from~(\ref{actgenfer}),
\begin{equation}
\left[\ga^0 \partial_0 + \ga^i \, N \, \partial_i + M \right]
X = 0 \,,\label{eomgen}
\end{equation}
and we multiply them on the left by $[ \ga^0
\partial_0 + \ga^i \, N \, \partial_i + M
]^\dagger$, we get
\begin{equation}
\Big\{\partial_0^2 - N^\dagger \, N \, \partial_i^2 + M^\dagger\,
M - \ga^0 \left[\left(\partial_0 N \right) \ga^i \,
\partial_i + \left( \partial_0 M \right) \right] \Big\} X
= 0 \,.
\end{equation}
We thus require $N^\dagger \, N \equiv \identity$. This
rewrites on the conditions
\begin{equation}
N_1^2 + N_2^2 = \identity \,,\qquad \left[N_1,\, N_2 \right]
= 0 \,.\label{condn}
\end{equation}

Our strategy is to reduce this problem to the one we have already
discussed. That is, we perform some redefinitions of the fields
to put the action~(\ref{actgenfer}) into the form~(\ref{act2fer}),
where we perform the canonical quantization of the system in the
way described in the previous subsection. The first of these
redefinitions strongly relies on the above
conditions~(\ref{condn}). If $N$ is a unitary matrix, we can find
a hermitean matrix $\Phi$ such that
\begin{equation}
N = \exp \left(2 \, \Phi \, \ga^0 \right),\qquad
\Phi^\dagger = \Phi \,.\label{defphi}
\end{equation}
Due to the properties of the $\ga^0$ matrix, this
amounts to
\begin{equation}
\cos \left(2 \, \Phi \right) = N_1 \,,\qquad
\sin \left(2 \, \Phi \right) = N_2 \,.
\end{equation}

After the redefinition $X \equiv \exp (- \ga^0 \,
\Phi ) {\hat X}$, the equations of
motions~(\ref{eomgen}) acquire the form
\begin{equation}
\left( \ga^0 \, \partial_0 + i \, \ga^i \, k_i
+ {\hat M} \right) {\hat X} = 0 \,,\label{eomhat}
\end{equation}
where we have expanded the fermions into plane waves $X_i (
\eta,\, \mathbf{k} ) =
\mathrm{e}^{i\,\mathbf{k \cdot x}} \, X_i ( \eta
)$ and introduced the new ``mass matrix''
\begin{eqnarray}
{\hat M} &=& \exp \left(\ga^0 \, \Phi \right) \left[ M
+ \ga^0 \partial_0 \right] \exp
\left( - \ga^0 \, \Phi \right)\nonumber\\
&\equiv& {\hat M}_1 + \ga^0 \, {\hat M}_2 \,.\label{masshat}
\end{eqnarray}
Notice that the two matrices ${\hat M}_1$ and ${\hat
M}_2$ are symmetric and antisymmetric, respectively.
This means that, in the one field case, one recovers the
standard equation
\begin{equation}
\left(\ga^0 \, \partial_0 + i \, \ga^i \, k_i + m \right)
{\hat \theta} = 0
\end{equation}
for spin $1/2$ fermions.

We have instead to perform a further redefinition of the
fields.\footnote{Contrarily to naive expectations, the
combination
\begin{equation}
\cos\; \Phi \: \partial_0 \: \cos \; \Phi + \sin \; \Phi \:
\partial_0 \:\sin \; \Phi \subset {\hat M}_2\label{probh}
\end{equation}
can be non vanishing at late times, even if the matrix
$N$ is approaching $\identity$. This occurs for example
in the application that we discuss in the next section.
If in that case we canonically defined the hamiltonian
$H$ starting with the fields ${\hat X}$, the
term~(\ref{probh}) would give $H$ a contribution
proportional to $\ga^0$ which does not vanish at late
times. The procedure described in the main text removes
this~problem.} Setting ${\hat X} = L \, \Xi$, we have
\begin{equation}
L^T  \left[ L \left( \ga^0 \, \partial_0 + i \,
\ga^i\, k_i \right) + {\hat M}_1 \, L + \ga^0
\left(\partial_0 + {\hat M}_2 \right) L
\right]  \Xi = 0\,.
\end{equation}
The matrix $L$ can be chosen such that $(\partial_0 + {\hat M}_2
) L = 0$. In particular, since ${\hat M}_2$ is antisymmetric and
real, $L$ can be taken orthogonal. The equations of motion for
the fields $\Xi$ can be thus cast in the form
\begin{equation}
\left( \ga^0 \, \partial_0 + i \, \ga^i \, k_i + L^T \,
{\hat M_1} \, L \right) \Xi = 0\,,
\end{equation}
that is with the identity matrix multiplying the term
which depends on the momentum and without any $\ga^0$
dependence in the ``mass'' matrix.

These equations (and the respective action for the
fields $\Xi_i$) are exactly of the form considered in
the previous subsection, so that we can apply the
quantization procedure discussed there.  As before, the
procedure is to canonically define the hamiltonian
starting from the set of fields $\Xi$ and to expand it
in a basis of creation/annihilation operators. The
occupation numbers are then calculated after the
diagonalization of the hamiltonian. It is possible to
show that this procedure can be carried out starting
from any of the basis for the fermionic fields, once the
hamiltonian has been canonically defined in the basis
$\Xi$. One can indeed explicitly verify that these
calculations lead to the same results for the occupation
numbers of the physical eigenstates, in the same way as
in the bosonic case the result~(\ref{efbos}) can be
computed starting from each of
expressions~(\ref{hambos}) or~(\ref{hambos2}).

We thus have
\begin{eqnarray}
H &\equiv& {\bar \Xi}  \left[ i \, \ga^i \, k_i + L^T \,
{\hat M_1} \, L \right]  \Xi = {\bar {\hat X}}  \left[
i \, \ga^i \, k_i + {\hat M_1} \right]  {\hat X}\nonumber\\
&=& {\bar X} \left[ i \, \ga^i \, k_i\, \mathrm{e}^{2\,\ga^0 \,
\Phi} + \mathrm{e}^{-\,\ga^0 \, \Phi} \, {\hat M}_1 \,
\mathrm{e}^{\ga^0 \, \Phi} \right]  X \,,\label{hamgen}
\end{eqnarray}
depending on which basis we work. In particular, when
working with the ${\hat X_i}$ or the $X_i$ fields, the
explicit knowledge of the matrix $L$ is not
needed.\footnote{However it is crucial that ${\hat M}_2$
is antisymmetric, which allows $L$ to be orthogonal.} We
present here the computation in the initial basis $X_i$,
which we found more convenient (in the numerical
calculations) for the application that we present in the
next section. In this basis, the
hamiltonian~(\ref{hamgen}) has the form
\begin{equation}
H = {\bar X}  \left[ i \, \ga^i \, k_i N + {\tilde M}_1 +
\ga^0 \, {\tilde M}_2 \right] X \,,\label{hamtilde}
\end{equation}
where the matrices ${\tilde M}_1$ and ${\tilde M}_2$ can
be obtained from eqs.~(\ref{hamgen}) and
(\ref{masshat}). At the end of the evolution, we simply
have ${\tilde M}_1 + \ga^0 \, {\tilde M}_2 = M_1$
diagonal, so that one recovers the ``standard''
hamiltonian for a system of $N$ decoupled fermions whose
masses coincide with the ones of the equations of motion
(which also become ``standard'').

To analyze the system during the evolution, we instead
decompose the spinors $X_i$ as in eqs.~(\ref{exp1ferm})
and~(\ref{spin}).\footnote{The charge conjugation matrix
now reads ${\tilde C} = - \ga^0 \, \ga^2$, so that
conditions~(\ref{majorana}) are replaced~by
\begin{equation}
V_+ = - i \, U_-^* \,,\qquad V_- = i \, U_+^* \,.
\end{equation}}
Taking the third coordinate along the momentum $k$, the
equations of motion~(\ref{eomgen}) rewrite:
\begin{equation}
\dot{U}_\pm = \mp \, i \left( M_1 \mp \, i \, M_2 \right) U_\pm
- i \, k \, \left( N_1 \pm \, i \, N_2 \right) U_\mp \,.
\label{eq:3.47}
\end{equation}
It is straightforward to check that they preserve the
conditions
\begin{equation}
U_+ \, U_+^\dagger + U_-^* \, U_-^T = 2 \cdot \identity\,,
\qquad U_+ \, U_-^+ = U_-^* \, U_+^T \,,\label{normalgen}
\end{equation}
which ensure the consistency of the \pagebreak[3] canonical
quantization.

We then expand the hamiltonian formally as in
eq.~(\ref{hdecferm}), where now the $\mathcal{E}$ and
$\mathcal{F}$ matrices read
\begin{eqnarray}\label{defef}
\mathcal{E} &=& U_+^\dagger \, k \left[ N_1 + i \, N_2
\right] U_- + U_-^\dagger \, k \left[ N_1 - i \, N_2 \right]
U_+ + \nonumber\\ &&+\, U_+^\dagger \left[ {\tilde M}_1
- i \, {\tilde M}_2 \right] U_+ + U_-^\dagger \left[ -
{\tilde M}_1 -  i \, {\tilde M}_2 \right] U_-,
\nonumber\\
\mathcal{F} &=& U_+^T \, k \left[ - N_2 - i \, N_1 \right]
U_+ + U_-^T \, k \left[- N_2 + i \,N_1 \right] U_- + \nonumber\\
&&+\, U_+^T \left[- {\tilde M}_2 + i \, {\tilde M}_1
\right] U_- + U_-^T \left[ - {\tilde M}_2 + i \, {\tilde M}_1
\right] U_+ \,.
\end{eqnarray}
We notice the properties
\begin{equation}
\mathcal{E}^\dagger = \mathcal{E} \,,\qquad
\mathcal{F}^T = \mathcal{F} \,.\label{relef}
\end{equation}

The matrix entering in eq.~(\ref{hdecferm}) can be
diagonalized through a unitary matrix~$C$
\begin{equation}
C \; \overline{H} \left( k, \eta \right) C^\dagger = H_d
\left( k, \eta \right) \qquad \mbox{diagonal} \,,\label{eqd1}
\end{equation}
such that the energy density is
\begin{equation}
\mathcal{H} = \left( a^+, b \right)\overline{H}
\pmatrix{a \cr b^\dagger}\equiv \left( {\hat a}^+, {\hat b}
\right) H_d \pmatrix{{\hat a} \cr {\hat b}^\dagger}.\label{eqd2}
\end{equation}

A first step in this diagonalization can be made by
noticing that the matrix $\overline{H}$ can be rewritten
as
\begin{equation}
\overline{H} = \mathcal{U}^\dagger \: \overline{H}_0 \:
\mathcal{U} \,,\label{defh0}
\end{equation}
with
\begin{eqnarray}
\overline{H}_0\equiv \pmatrix{- {\tilde M}_1 + i\,{\tilde M}_2
& k\,N_1 + i\,k\,N_2 \cr k\,N_1 - i\,k\,N_2 & {\tilde
M}_1 + i\,{\tilde M}_2}
\end{eqnarray}
hermitean and
\begin{eqnarray}
\mathcal{U} \equiv \frac{1}{\sqrt{2}}
\pmatrix{U_+ & -i\,U_-^*\cr - U_- & -i\,U_+^*}
\end{eqnarray}
unitary, as it follows from eqs.~(\ref{normalgen}).

We are not able to provide general analytical formulae
for the diagonalization of the remaining matrix
$\overline{H}_0$. This diagonalization can however be
performed numerically. In addition, some important
conclusions can be drawn from the properties of the
matrix $\mathcal{H}$. Due to the
relations~(\ref{relef}), one can show (i.e.\ by counting
the number of independent equations that must be
satisfied) that the matrix $C$ entering in
eq.~(\ref{eqd1}) can be of the form
\begin{eqnarray}
C \equiv \pmatrix{I & J \cr i \, J^* & -  i \, I^*}
\label{defc}
\end{eqnarray}
(where $I$ and $J$ are $N \times N$ matrices). Unitarity
of $C$ requires
\begin{equation}
I \, I^\dagger + J \, J^\dagger = \identity \,,\qquad
I^\dagger \, J = J^T \, I^* \,.\label{unitarity}
\end{equation}

By explicitly performing the product~(\ref{eqd1}) one
realizes that the eigenvalues of the hamiltonian occur
in pairs, and that $H_d$ is of the form \\
$\diag (\omega_1,\, \omega_2,\, \dots,\, \omega_N,-
\omega_1,- \omega_2,\,\dots \,, -  \omega_N) $.
The eigenstates corresponding to each couple $\pm
\,\omega_i$ are to be interpreted as particle and
antiparticle states with the same energy. Finally, it is
possible to show that particles and antiparticles are
produced in the same amount. Defining the vacuum state
to be annihilated by the initial annihilation operators
$a_i$ and $b_i$, we have indeed (we remind that in this
expression we do not sum over $i$)
\begin{equation}
N_i \left( \eta \right) = \langle {\hat a}_i^\dagger \,
{\hat a}_i \rangle = \langle {\hat b}_i^\dagger \, {\hat
b}_i \rangle = \left( J \, J^\dagger \right)_{ii}.
\label{occnum}
\end{equation}
We assume that no fermionic particles are present at initial time
$\overline{\eta} $. In our formalism, this is equivalent to
requiring $J ( \overline{\eta} ) = 0$. Notice also that the
unitarity condition~(\ref{unitarity}) ensures that the Pauli
principle is always fulfilled.

\section{One application: non-thermal gravitino production}
\label{gravi}

In this section we discuss one application of the above
formalism, i.e.\ non-thermal gravitino production in a
system with two chiral superfields. This model consists
of two chiral multiplets coupled only gravitationally.
The scalar of the first multiplet is responsible for
driving inflation, while the one of the second breaks
supersymmetry in the vacuum. The motivations for this
analysis, as well for the specific model chosen, are
explained in the introduction (see also~\cite{noi}).

This section is divided in five subsections. In the first one we
introduce all the quantities relevant for this calculation.
Although we have exactly followed the conventions of
ref.~\cite{kklv2}, the aim of section~\ref{sub1} is to provide a
practical self contained presentation. In section~\ref{sub2} we
describe the model that we are considering. We also discuss there
the evolution of the scalar fields, which constitute the external
background for the fermionic fields. In section~\ref{sub3} we
show how to apply the formalism of the previous section to the
calculation of the abundances of the fermions of the theory. Our
results are presented in the two remaining subsections. In
section~\ref{sub4} we present analytical results in the case in
which supersymmetry is actually unbroken in the vacuum of the
theory. We show that in this case, gravitinos are only
gravitationally (hence negligibly) produced. Already this
consideration suggests that in the class of models we are
considering (i.e.\ with the two sectors  coupled only
gravitationally) non-thermal gravitino production may be very
inefficient in the realistic situation in which the observable
supersymmetry breaking (TeV scale) is much smaller than the scale
of inflation ($10^{13}\mathrm{GeV}$). This is confirmed by the
numerical results presented in section~\ref{sub5}, which show
that gravitino production indeed decreases as the size of
supersymmetry breakdown becomes smaller.

\subsection{Definitions} \label{sub1}

We write here the relevant equations of motion for the
gravitino field and the fermionic particles to which it
is coupled. We follow the conventions of
ref.~\cite{kklv2}. The starting action is the one of
$D=4, N=1$ supergravity, with four fermion
interactions omitted. For simplicity, we do not consider
any gauge multiplet, but for the moment we allow chiral
superfields to be complex. The lagrangian reads
\begin{eqnarray}
e^{-1} \mathcal{L} &=& -\frac12 \, \mpl^2 \, R - g_i{}^j
\left( \partial_\mu \, \phi^i \right) \left(
\partial^\mu \phi_j \right) -  V -\nonumber\\ &&-\,
\frac12 \, \mpl^2 \, \bar \psi_\mu \, R^\mu + \frac12 \,
m \, \bar \psi_{\mu R} \, \gamma^{\mu \nu} \, \psi_{\nu
R} + \frac12 \, m^* \, \bar \psi_{\mu L} \, \gamma^{\mu
\nu } \, \psi_{\nu L}- \nonumber\\ &&-\, g_i{}^j \left[
\bar \chi_j \, \Dbar \chi^i + \bar \chi^i \Dbar \chi_j
\right] - m^{ij} \, \bar \chi_i \, \chi_j - m_{ij} \,
\bar \chi^i \, \chi^j +\nonumber\\ &&+\, \Big( 2 \,
g_j{}^i \bar \psi_{\mu R} \, \gamma^{\nu \mu} \, \chi^j
\, \partial_\nu \phi_i + \bar \psi_R \cdot \gamma
\upsilon_L + \mbox{h.c.} \Big)\,.\label{lagin}
\end{eqnarray}

The first line of eq.~(\ref{lagin}) concerns the scalar fields.
The first term is the standard one of Einstein gravity, with
$\mpl$ denoting the reduced Planck mass ($\mpl \simeq 2.4 \cdot
10^{18}\,\mathrm{GeV}$) and $R$ the Ricci scalar. Conformal time
$\eta$ is used and the Minkowski metric is taken with signature
$-+++$. More explicitly, the metric and the vierbein are given by
$g_{\mu \, \nu} = a^2 ( \eta )\eta_{\mu \,
\nu}, e_\mu^b = a ( \eta ) \delta_\mu^b,$
where $a$ is the scale factor of the Universe. We then have some
chiral complex multiplets formed by $(
\phi_i,\, \chi_i )$ and their conjugate $(
\phi^i,\, \chi^i )$. It is worth emphasizing that
$\chi_i$ is a left  handed field, while $\chi^i$ a right
handed one. The left and right projections are $P_L
\equiv ( 1 + \gamma_5
)/2, P_R \equiv ( 1 - \gamma_5 )/2$. The gamma matrices in curved
space $\gamma$ are related to the ones in flat space $\ga$ by the
relation $\gamma^\mu = a^{- 1} \, \ga^\mu$, and the realization
of the latter that we are using is given in eq.~(\ref{gammas}).
The K\"ahler metric is the second derivative of the K\"ahler
potential
\begin{equation}
g_j{}^i = \frac{\partial}{\partial \, \phi_i} \,
\frac{\partial}{\partial \, \phi^j} \: K\,.
\end{equation}
while the scalar potential $V$ is defined below.

In the second line of eq.~(\ref{lagin}) we have the
kinetic and the mass term for the gravitino field. The
first one is defined to be
\begin{equation}
R^\mu = e^{-1} \, \epsilon^{\mu \nu \rho \sigma} \,
\gamma_5 \, \gamma_\nu \, D_\rho \, \psi_\sigma\,,
\end{equation}
where the covariant derivative
\begin{equation}
D_\mu \psi_\nu = \left( \left( \partial_\mu +
\frac{1}{4} \omega_\mu^{m n} \gamma_{m n} \right)
\delta_\nu^\lambda - \Gamma_{\mu \, \nu}^\lambda \right)
\psi_\lambda\,,
\end{equation}
contains the spin connection $\omega_\mu^{m\,n}$ and the
connection $\Gamma_{\mu \, \nu}^\lambda$ ($\gamma_{m\,n}
\equiv [ \ga_m, \ga_n ] /2$).
The mass parameter $m$ is instead given by
\begin{equation}
m \equiv \mathrm{e}^{\frac{K}{2\,\mpl^2}} \: W\,,
\end{equation}
and it is related to the gravitino mass by
\begin{equation}
m_{\tilde G} = \vert m \vert \: \mpl^{-2}\,.
\end{equation}

We then find the kinetic and mass term for the chiral
fermions. The first is given~by
\begin{equation}
D_\mu \chi_i \equiv \left( \partial_\mu + \frac{1}{4}
\omega_\mu^{m n} \, \gamma_{m n} \right) \chi_i +
\frac{1}{4 \, \mpl^2} \left[ \partial_j K \,
\partial_\mu \phi^j - \partial^j K \, \partial_\mu
\phi_j \right] \chi_i + \Gamma_i^{j\,k} \chi_j
\partial_\mu \phi_k\,.
\end{equation}
where $\Gamma_i^{j\,k} \equiv  {g^{-1}}_i{}^l \,
\partial^j \, g_l{}^k$ is the K\"ahler connection.
For what concerns instead the fermion masses, we have
\begin{eqnarray}
m^i &\equiv& D^i m \equiv \partial^i m + \frac{1}{2\,\mpl^2}
\partial^i K \, m\,, \nonumber\\
m^{i j} &\equiv& D^i \, D^j m = \left( \partial^i +
\frac{1}{2\,\mpl^2}\partial^j K \right) m^j - \Gamma_k^{i\,j} \,
m^k\,.
\end{eqnarray}

\noindent We can now write, in compact notation, the scalar
potential
\begin{equation}\label{scalpot}
V \equiv -  3 \, \mpl^{-2} \vert m \vert^2 + m_i \,
{g^{-1}}_j{}^i  \, m^j\,.
\end{equation}

The last line of eq.~(\ref{lagin}) describes the
interactions of the gravitino with the chiral fields
(i.e.\ with matter). The field $\upsilon_L$ is defined
to be
\begin{equation}
\upsilon_L \equiv m^i \, \chi_i + \left( \dbar \phi_i \right)
\chi^j \, g_j{}^i\,.\label{gold}
\end{equation}
As understood in refs.~\cite{grt2,kklv2}, this combination of
matter fields is the goldstino (actually its left-handed
component) in a cosmological context, where supersymmetry is
broken both by the kinetic and the potential energies of the
scalar fields. We work in the unitary gauge, where the goldstino
is gauged away to zero. We also Fourier transform the fermion
fields in the spatial direction, i.e.\ $\chi ( \eta,
\vec{x} ) \equiv \chi ( \eta )
\mathrm{e}^{i \, x^i \, k_i}$.

The gravitino field has transversal and longitudinal
components. To appreciate their different behavior, one
can introduce the projectors~\cite{kklv2}
\begin{eqnarray}
\left( P_\gamma \right)_i &\equiv& \frac{1}{2}
\left(\ga^i - \frac{1}{\vec{k}^2} \, k_i \left( k_j \, \ga^j
\right)\right), \nonumber\\
\left(P_k \right)_i &\equiv& \frac{1}{2 \, \vec{k}^2}
\left( 3 \, k_i - \ga_i  \left( k_j \, \ga^j \right) \right),
\end{eqnarray}
where $k_i$ are the spatial components of the comoving
momentum of the gravitino (i.e.\ $\partial_0 k_i = 0$)
and $\vec{k}^2 \equiv k_i \, k_i$. These projectors are
employed in the decomposition
\begin{equation}\label{deco}
\psi_i = \pt + \left( P_\gamma \right)_i \, \ph +
\left(P_k\right)_i \, \pk\,,
\end{equation}
where $\ph \equiv \ga^i \, \psi_i$.\footnote{Notice that
\begin{eqnarray}
k_i \left( P_\gamma \right)_i &=& 0\,, \qquad \ga^i
\left(P_\gamma \right)_i = 1\,, \nonumber\\
k_i  \left(P_k \right)_i &=& 1\,, \qquad \ga^i
\left(P_k \right)_i = 0\,, \nonumber\\
k_i \psi_i^T &=& \ga^i \psi_i^T = 0\,.\label{proi1}
\end{eqnarray}}
As it is shown in ref.~\cite{kklv2}, from the
lagrangian~(\ref{lagin}) one recovers four independent
equations for the gravitino components. Two of them are
algebraic constraints which involve $\psi_0,\: \pk,$ and
$\ph$. We use them to eliminate the first two
combinations in favor of the last one. The other two are
instead dynamical, and can be written in the form
\begin{eqnarray}
\left[ \ga^0 \, \partial_0 + i \, \ga^i \, k_i +
\frac{\da \, \ga^0}{2} + \frac{a \, \underline{m}}{\mpl^2} \right]
\pt &=& 0\,, \label{eqt} \\
\left(\partial_0 + \hat{B} + i \, \ga^i \, k_i \, \ga^0 \,
\hat{A} \right) \ph - \frac{4}{\alpha \, a} \, k^2 \,
\Upsilon &=& 0\,, \label{eqph}
\end{eqnarray}
where
\begin{eqnarray}
\Upsilon &=& g_j{}^i \left( \chi_i \, \partial_0 \, \phi^j
+ \chi^j \, \partial_0 \, \phi_i \right),\nonumber\\
\underline{m} &=& P_R \, m + P_L \, m^*\,,\qquad \vert m \vert^2 =
\underline{m}^\dagger \, \underline{m}\,, \nonumber\\
\hat{A} &=& \frac{1}{\alpha} \left( \alpha_1 - \ga^0
\, \alpha_2 \right),\qquad \hat{B} = - \frac{3}{2} \,
\da \, \hat{A} + \frac{1}{2 \, \mpl^2} \, a \,
\underline{m} \, \ga^0 \left(1 + 3 \, \hat{A} \right),\nonumber\\
\alpha &=& 3 \, \mpl^2\left(H^2+ \frac{\vert m \vert^2}{\mpl^4}
\right),\nonumber\\ \alpha_1 &=& - \mpl^2  \left(3 \, H^2
+ 2 \, \dot{H} \right) - \frac{3}{\mpl^2} \, \vert m
\vert^2\,,\qquad \alpha_2 = 2 \, a^{-1} \, \partial_0
\underline{m}^\dagger\,,\label{manydef}
\end{eqnarray}
$\dot{f} \equiv a^{-1} \partial_0 f$, and where $H
\equiv \da / a$ is the Hubble expansion rate.

We notice that the transverse component of the
gravitino, $\psi_i^T$, is decoupled from the
longitudinal component and from matter, apart from
gravitational effects due to the expanding background.
In particular, transverse gravitinos are produced only
gravitationally~\cite{kklv1,grt1,mar}, and for this
reason we will not consider this component any longer in
the rest of the work.

We are thus left with the gravitino longitudinal
component, rewritten in terms of $\ph$, and the matter
fields. In case of only one chiral supermultiplet the
combination $\Upsilon$ defined above is proportional to
the goldstino, and thus vanishes in the unitarity gauge.
In the more general case of $N$ chiral superfields, we
have (always in the unitary gauge) $N-1$ non vanishing
independent fermionic chiral fields, and one should go
into a basis orthogonal to the goldstino. The equations
of motion for all these fields can be of course deduced
from the initial lagrangian~(\ref{lagin}). If only two
superfields are present, one is just left with the
matter field $\Upsilon$ defined above.

\subsection{Description of the model considered and evolution\\
of the scalar fields} \label{sub2}

The matter content of the model we are considering is of
two superfields $\Phi$ and $S$, with superpotential
\begin{equation}
W=\frac{m_\phi}{2}\,\Phi^2+\mu^2\left(\beta+S\right)
\label{superpot}
\end{equation}
and minimal K\"ahler potential
\begin{equation}
K=\Phi^\dagger\,\Phi+S^\dagger\, S\,.
\end{equation}

The potential for the scalar components $\phi$ and $s$
of the superfields $\Phi$ and $S$ can be computed using
eq.~(\ref{scalpot}). We then assume that the scalar
fields are real, that is (after $V$ is computed) we
perform the substitutions
\begin{equation}
\phi=\phi^*\longrightarrow \frac{\phi}{\sqrt{2}}\,,\qquad
s=s^* \longrightarrow \frac{s}{\sqrt{2}}\,.
\end{equation}
In this way the real scalar fields have canonical
kinetic terms.

During inflation, the field $\phi$ acts as the inflaton, while
the v.e.v.\ of $s$ is quickly driven to $\langle s
\rangle \simeq 0$. The potential is then practically
the one of chaotic inflation, and $m_\phi \sim
10^{13}\,\mathrm{GeV}$ must be posed to match the COBE
results for the size of the
CMBR~fluctuations.\footnote{As it is known, the
contributions from the K\"ahler potential to the scalar
potential are very relevant for $\phi\sim \mpl$. This is
a common problem for supersymmetric theories of
inflation, where the $F$ terms generically spoil the
flatness of the potential during the inflationary regime
(for a review, see~\cite{lyri}; see also~\cite{buc} for
a recent discussion). As a consequence, the theory that
we are here describing should be modified during
inflation; however, we will not consider this issue here
and we will still assume that the value for $m_\phi$ is
not too different from the one imposed in ``usual''
chaotic~inflation.}

At the end of inflation, the field $\phi$ oscillates about the
minimum $\phi = 0$. The amplitude of these oscillations is dumped
by the expansion of the Universe (and, later on, also by the
decay of the inflaton that every realistic model must include).
If only $\phi$ was present, we eventually would have unbroken
supersymmetry in the minimum $\phi=0$. The role of the $s$ field
is to provide the supersymmetry breaking in the minimum. The
second term in eq.~(\ref{superpot}) is known as the Polonyi
superpotential~\cite{polo}, and it provides a simple example on
how supersymmetry can be broken in a hidden sector and
transmitted to the observable one by gravity. What is remarkable
with this potential is that, for particular values of the
parameter $\beta$, supersymmetry is broken with a vanishing value
for the cosmological constant. If indeed we take $\beta= ( 2 -
\sqrt{3} ) \mpl$ (for a more detailed discussion, see for
example~\cite{tm}), the potential $V ( \phi = 0,\: s )$ vanishes
in its minimum at
\begin{equation}\label{min}
s_0 = \sqrt{2} ( \sqrt{3} - 1 ) \mpl\,.
\end{equation}

For this value, the gravitino mass $m = \mathrm{e}^{K /
2 \, \mpl^2} \: W$ evaluates to
\begin{equation}\label{mgra}
m_{\tilde G}= \mathrm{e}^{2-\sqrt{3}} \:
\frac{\mu^2}{\mpl}  \simeq 1.31 \: \frac{\mu^2}{\mpl}\,,
\end{equation}
which is a typical result for this breaking of
supersymmetry. We see that the ``intermediate'' scale
$\mu$ must be taken of order $10^{10}\mathrm{GeV}$ to
reproduce the expected gravitino mass $m_{\tilde G} \sim
100\,\mathrm{GeV}$.

In the following, we discuss in more details the
evolution of the two scalar fields. To do this, we use
physical time $t$ and work with the adimensional
quantities
\begin{eqnarray}
{\hat \phi} &\equiv& \frac{\phi}{\mpl}\,,\qquad {\hat s} \equiv
\frac{s}{\mpl}\,,\qquad {\hat \beta} \equiv \frac{\beta}{\mpl}\,,
\nonumber\\
{\hat t} &\equiv& t \, m_\phi\,,\qquad {\hat \mu^2} \equiv
\frac{\mu^2}{\mpl \, m_\phi}\,,\qquad {\hat H} \equiv \;\;
\frac{H}{m_\phi}\,,\qquad {\hat V} \equiv \;\; \frac{V}{\mpl^2 \,
m_\phi^2}\,,\label{adim}
\end{eqnarray}
where we remind that $H$ and $V$ are, respectively, the
Hubble constant and the scalar potential. In terms of
these redefined quantities, the equations of motion for
the two scalars read
\begin{equation}
\frac{d^2 {\hat \phi}_i}{d {\hat t}^2} + 3 \, {\hat H} \,
\frac{d {\hat \phi}_i}{d {\hat t}} +
\frac{d {\hat V}}{d {\hat \phi}_i} = 0\,,\qquad
{\hat \phi}_i = {\hat \phi},\, {\hat s}.
\end{equation}

\FIGURE[t]{\epsfig{file=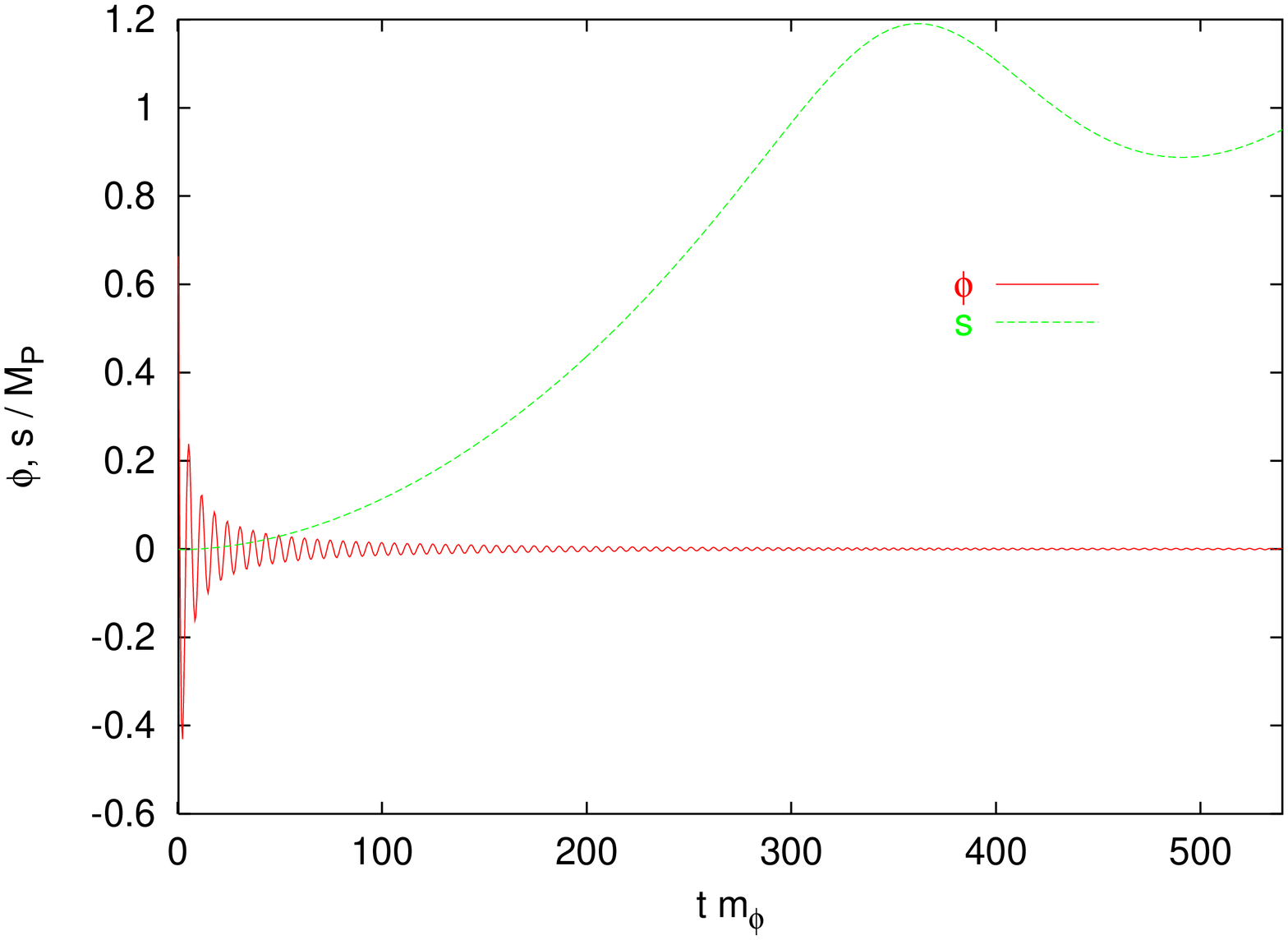, angle=-90, width=0.8\textwidth}
\caption{Evolution of the two scalar fields $\phi$ and $s$ for
${\hat \mu}^2 = 10^{-2}$.} \label{fig1}}

We start our numerical calculations at ${\hat \phi}
\simeq 1.4$, short after inflation, and with the scale
factor $a$ normalized to one.\pagebreak[3]

We show in figure~\ref{fig1} the evolution for the two
scalar fields after inflation, in the case ${\hat \mu}^2
= 10^{-2}$. As we said, initially the model reproduces
the scalar potential of chaotic inflation, and thus we
have
\begin{equation}
{\hat \phi} \simeq \sqrt{\frac{8}{3}} \, \frac{\cos\;
{\hat t}}{{\hat t}}\,,\qquad {\hat s} \simeq 0\,.\label{phiin}
\end{equation}
The initial dynamics of the Polonyi field $s$ is determined by
$\phi$. More precisely, we can write an effective potential $V (
s )$ for it by substituting eq.~(\ref{phiin}) into $V (\phi,\, s
)$ and then averaging over the inflaton oscillations. Expanding
$V$ for both ${\hat \phi}$ and ${\hat s}$ smaller than one, we
find that the potential is minimized by
\begin{equation}
{\hat s}_{\min} \simeq \frac{\sqrt{2} \, {\hat \mu}^2 [ 16 \,
{\hat \beta} \, {\hat \mu}^2 - \langle {\hat \phi}^4 \rangle ]}{4
\, \langle {\hat
\phi}^2 \rangle - 16 \, {\hat \beta^2} \, {\hat \mu}^4}
\simeq \frac{3 \, \sqrt{2} \, {\hat \beta} \left(
{\hat \mu}^2 \, {\hat t} \right)^2}{1 - 3 \, {\hat
\beta}^2 \left( {\hat \mu}^2 \, {\hat t} \right)}\,.\label{pilin}
\end{equation}
To be precise, the Polonyi field is always smaller than ${\hat
s}_{\min}$, due to the fact that the expansion of the Universe
slows its motion towards the minimum of $V (s)$. However,
eq.~(\ref{pilin}) gives a good estimate for the order of
magnitude of $s$ in this initial stage.

What is most important to emphasize, is that
eqs.~(\ref{phiin}) and~(\ref{pilin}) explicitly show the
presence of two very different (physical) time-scales in
the model we are considering. The first of them is set
by the inverse inflaton mass $m_\phi^{-1}$ which is
the time-scale of the oscillations of the inflaton
field. The second one is given by ${\hat\mu^2} \,
m_\phi^{-1}$. Equation~(\ref{pilin}) shows that this
is the relevant time-scale for the Polonyi field in the
initial stage. However, this is true also for the
complete evolution of $s$. To see this, let us consider
the latest times shown in figure~\ref{fig1}. In this
stage the amplitude of the oscillations of $\phi$ are
negligible. The evolution of $s$ is not any longer
influenced by the inflaton field, but it starts
oscillating about the minimum of its own potential given
in eq.~(\ref{min}).\footnote{There is of course a
possible moduli problem associated with these
oscillations. However, we do not consider this issue
here.} The amplitude of these oscillations is also
dumped by the expansion of the Universe, while their
period is related to the inverse Polonyi mass, which is
now (i.e.\ at $\phi=0$) given by~\cite{tm}
\begin{equation}
m_s = \sqrt{2 \, \sqrt{3}} \, m_{\tilde G} \simeq 2.4 \,
{\hat \mu}^2 \, m_\phi\,.
\end{equation}
The quantity ${\hat \mu}^2$ defines the ratio between the two
scales. In figure~\ref{fig1} we have chosen, for illustrative
purposes, ${\hat \mu}^2 = 10^{-2}$. However, this value is
unphysical, since it would correspond to a too high supersymmetry
breaking scale. Indeed, as eq.~(\ref{mgra}) shows, we must require
${\hat \mu}^2 \sim 10^{-11}$, if supersymmetry is supposed to
solve the \mbox{hierarchy problem.}

While the size of ${\hat \mu^2}$ controls the
supersymmetry breaking in the vacuum of the theory, both
scalar fields contribute to break supersymmetry during
their evolution. \pagebreak[3] In particular, both their kinetic and
potential energies contribute to the breaking, as
emphasized in ref.~\cite{grt2}. This can be seen by
looking to the transformation law of the chiral fermions
$\chi_i$ under an infinitesimal supersymmetry
transformation with parameter $\varepsilon$. In our case
they read
\begin{equation}
\delta \chi_i = - \frac{1}{2} \, P_L \,
\left[ m_i - \frac{1}{\sqrt{2}} \, \ga^0 \, \frac{d \phi_i}{d t}
\right] \varepsilon\,,
\end{equation}
where $\phi_1 = \phi,\ \phi_2 = s$.

Following refs.~\cite{grt2,kklv2}, we define the
quantities
\begin{equation}
f_{\phi_i}^2 \equiv m_i^2 + \frac{1}{2} \left(
\frac{d \phi_i}{d t} \right)^2,
\end{equation}
which give a ``measure'' of the size of the
supersymmetry breaking provided by the $F$  term
associated with the $i$-th scalar field. More precisely,
we will be interested in the normalized quantities
\begin{equation}
r_\phi \equiv \frac{f_1^2}{f_1^2 + f_2^2}\,,\qquad r_s \equiv
\frac{f_2^2}{f_1^2 + f_2^2}\,,\label{defrs}
\end{equation}
which indicate the relative contribution of the two
scalar fields $\phi$ and $s$.

\FIGURE[t]{\epsfig{file=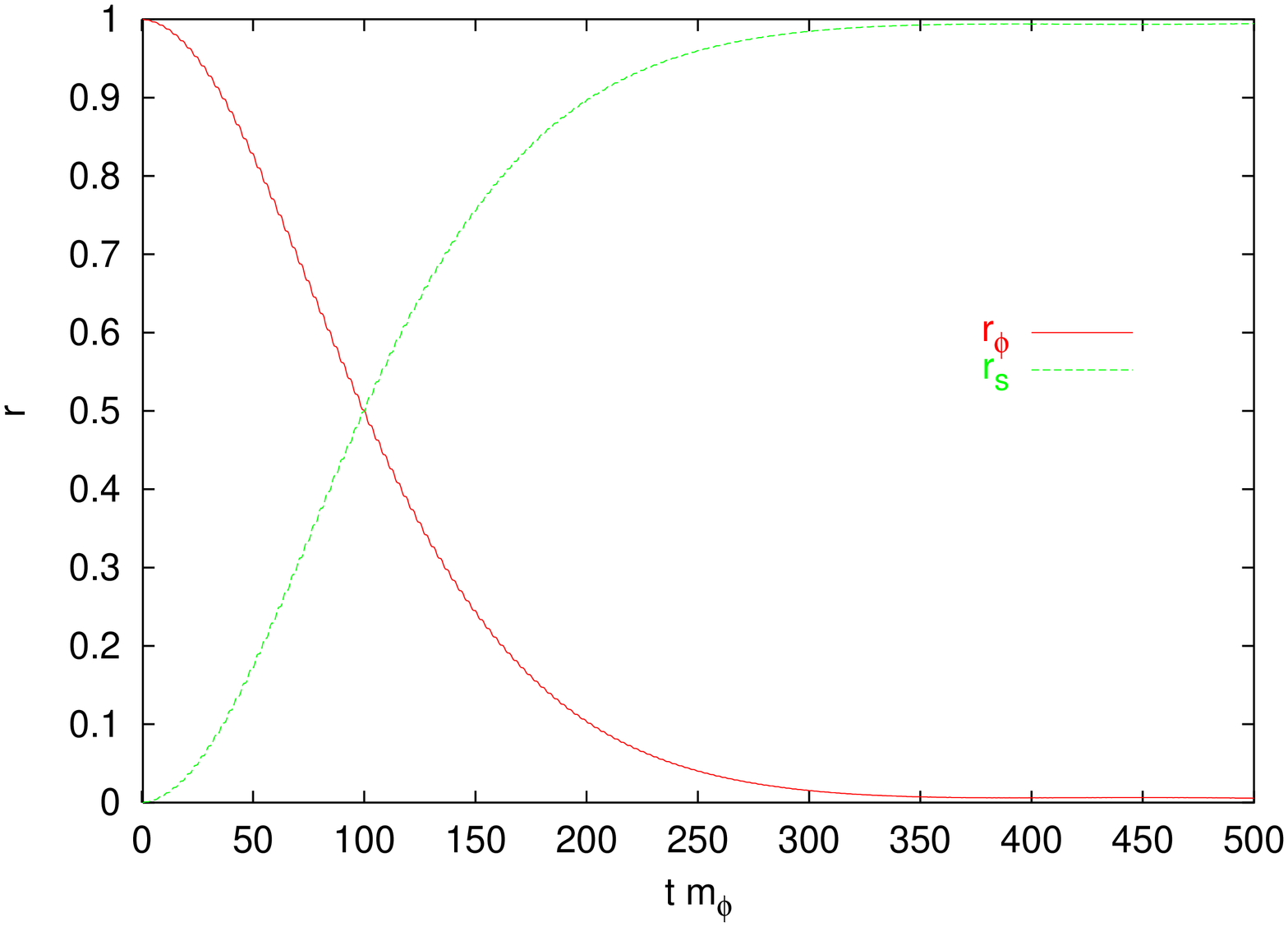, angle=-90, width=0.8\textwidth}
\caption{Relative contribution of the two scalar fields $\phi$
and $s$ to the supersymmetry breaking during their
evolution. As in figure~\ref{fig1}, ${\hat \mu}^2 =
10^{-2}$.}
\label{fig2}}

In figure~\ref{fig2} we have shown the evolution of
$r_\phi$ and $r_s$ for the specific case ${\hat \mu}^2 =
10^{-2}$. As expected, in the initial stages only
the inflaton contributes to the supersymmetry breaking,
while only the Polonyi contributes at later times. The
regime of equal contribution is around ${\hat t}={\hat
\mu}^{-2}$, when $\phi$ and $s$ are of the same size
(cf.\ figure~\ref{fig1}). As it should be clear from the
above discussion, $r_\phi$ and $r_s$ share the identical
behavior for all the choices of ${\hat \mu}^2$, once
${\hat t}$ is given in units of ${\hat \mu}^{-2}$.

\subsection{Effective fermionic lagrangian and hamiltonian in
the case\\ of two chiral supermultiplets} \label{sub3}

The fermionic content of the model we are considering is of the
gravitino $\psi_\mu$ and the two chiral fermions ${\tilde \phi}$
and ${\tilde s}$. In the unitary gauge, one combination of
${\tilde \phi}$ and ${\tilde s}$, the goldstino $\upsilon$, is
set to zero, while the transverse component of the gravitino,
$\pt$, is only gravitationally coupled to the other fields. The
other two fermions $\ph$ (the longitudinal gravitino component)
and $\Upsilon$ (the combination of chiral fermions orthogonal to
$\upsilon$) are coupled together, as we described in
section~\ref{sub1}.

With some algebra, we can rewrite the initial
lagrangian~(\ref{lagin}) in three terms
\begin{equation}
\mathcal{L} = \mathcal{L}_{\mathrm{background}} +
\mathcal{L}_{\pt}+ \mathcal{L}_{\ph \,\Upsilon}\,.
\end{equation}
The first term governs the dynamics of the scalar fields
and of the scale factor of the Universe. The second
describes the (decoupled) transverse gravitino
component, while the third one reads
\begin{eqnarray}
\mathcal{L}_{\ph \,\Upsilon} &=& - \frac{\alpha}{4 \, k^2} \,
a^3 \, \overline{\ph} \Bigg[ \ga^0 \, \partial_0 \ph + i \, \ga^i
\, k_i \hat{A} \, \ph  + \nonumber\\ &&
\hphantom{- \frac{\alpha}{4 \, k^2} \,
a^3 \, \overline{\ph} \Bigg[}\! - \left( \frac{3}{2} \, \da \,
\ga^0 +
\frac{3}{2 \, \mpl^2} \, a \, m \right) \hat{A} \, \ph -
\frac{a \, m}{2 \, \mpl^2} \, \ph - \frac{4 \,
k^2}{\alpha \, a} \, \ga^0 \, \Upsilon \Bigg] +
\nonumber\\ && - \, \frac{4 \, a}{\alpha \, \Delta^2} \,
\overline{\Upsilon} \Bigg[ \ga^0 \, \partial_0
\Upsilon - i \, \ga^i \, k_i \hat{A} \, \Upsilon
- \frac{3}{2} \, \da \, \ga^0
\, \hat{A} \, \Upsilon + \frac{3}{2 \, \mpl^2} \,
\hat{A} \, a \, m \, \Upsilon + \nonumber\\
&&\hphantom{- \, \frac{4 \, a}{\alpha \, \Delta^2} \,
\overline{\Upsilon} \Bigg[}\!+\,2 \, \da \, \ga^0 \,
\Upsilon - \frac{a \, m}{2 \, \mpl^2} \, \Upsilon +
\frac{1}{4} \, a \, \alpha \, \Delta^2 \, \ga^0 \, \ph
\Bigg]\,. \label{lagnoi}
\end{eqnarray}
We have expanded the fermions into plane waves $X_i (\eta,\,
\mathbf{k} ) = \mathrm{e}^{i\,{k_i x^i}} \, X_i ( \eta
)$, where $k_i$ is the comoving momentum (i.e.\ $\partial_0
\, k_i = 0)$, and we have introduced
\begin{eqnarray}
\Delta &\equiv& \frac{2}{\alpha} \left[ \dot{\phi}_i \,
\dot{\phi_j} \, m_k m_l \left( {g^{-1}}_{k l} \, g_{i j}
- \delta_{i k} \, \delta_{l j} \right) \right]^{1/2}
\nonumber\\ &=&\frac{2}{\alpha} \left( m_1
\dot{\phi}_2 - m_2 \dot{\phi}_1 \right),\label{defdel}
\end{eqnarray}
where the second equality holds in the case of a minimal
K\"ahler potential, $g_i^j = \delta_i^j$. The quantity
$\Delta$ has no counterpart in the one chiral superfield
case, and indeed it is negligible unless both the scalar
fields give a sizeable contribution to the breaking
of~supersymmetry.

One can explicitly verify that the
lagrangian~(\ref{lagnoi}) reproduces the equation of
motion~(\ref{eqph}) for the longitudinal gravitino
component, as well as the one for $\Upsilon$ that one
obtains from the initial lagrangian~(\ref{lagin}).
However, we notice that the two fields $\ph$ and
$\Upsilon$ are not canonically normalized. Canonical
normalization has to be imposed, if we want our fields
to give invariant quantities (as for example the
occupation number) in comoving units in the adiabatic
regime. Among the possible redefinitions, we choose
\begin{eqnarray}
\ph &=& \frac{2 \, i \, \ga^i \, k_i}{\left(\alpha \, a^3
\right)^{1/2}} \, {\tilde \ph}\,, \nonumber\\
\Upsilon &=& \frac{\Delta}{2} \left( \frac{\alpha}{a}
\right)^{1/2}  {\tilde \Upsilon}\,,\label{ridef1}
\end{eqnarray}
since the equations of motion look quite symmetric in
terms of the new fields. In matrix form, they are
exactly of the form~(\ref{eomgen}), i.e.\
\begin{equation}
\left( \ga^0 \, \partial_0 + i \, \ga^i \, k_i N + M
\right) X = 0\,,\label{eomtilde}
\end{equation}
where $X$ is the vector $( {\tilde \ph}, {\tilde
\Upsilon} )^T$. In our specific case, the ``mass''
matrix $M$ is given~by
\begin{eqnarray}
M &=& \diag  \left( \frac{m \, a}{2 \, \mpl^2} + \,
\frac{3}{2} \, \left( \frac{m \, a}{\mpl^2} \, \al1 +
\dot{a} \, \al2 \right) \right.- \nonumber\\
&&\hphantom{\diag  \Bigg(}\!\left. - \, \frac{m \, a}{2
\, \mpl^2} + \,\frac{3}{2} \left( \frac{m \, a}{\mpl^2}
\, \al1 + \dot{a} \, \al2 \right) + a  \left( m_{11} + m_{22}
\right) \right),\label{masstilde}
\end{eqnarray}
and the $N$ matrix by
\begin{equation}
N \equiv N_1 + \ga^0 \, N_2 = \pmatrix{- \al1 & 0 \cr 0
& - \al1} + \ga^0 \pmatrix{- \al2 & - \Delta
\cr -  \Delta & \al2}.
\label{kaptilde}
\end{equation}

In the above equations, we have defined ${\tilde
\alpha}_i \equiv \alpha_i / \alpha$. The relation
$\al1^2 + \al2^2 \equiv 1$ which holds in the one chiral
field case~\cite{kklv1,grt1} is now replaced
by\footnote{When only one scalar field gives a
substantial contribution to supersymmetry breaking, the
quantity $\Delta$ almost vanishes, and the relation
$\al1^2 + \al2^2 \simeq 1$ holds approximatively.}
\begin{equation}
\al1^2 + \al2^2 + \Delta^2 \equiv 1\,.\label{reldel}
\end{equation}
We thus see that the matrices $N_1$ and $N_2$ satisfy
both conditions~(\ref{condn}).

The equations of motion~(\ref{eomtilde}) have a clear
behavior in the low energy limit, when the two scalars
of the theory settle to their minima. In this final
stage one has $\al1 = -1, \al2 = \Delta = 0$, as it
can be easily checked from the definitions listed above.
As a consequence, eqs.~(\ref{eomtilde}) decouple, and
each of them acquires the standard form for spin $1/2$
fermions
\begin{eqnarray}
\left( \ga^0 \, \partial_0 + i \, \ga^i \, k_i +
a \, m_{\tilde \ph} \right) {\tilde \ph} &=&  0\,, \nonumber\\
\left( \ga^0 \, \partial_0 + i \, \ga^i \, k_i + a \,
m_{\tilde \Upsilon} \right) {\tilde \Upsilon} &=& 0\,,
\label{eomstand}
\end{eqnarray}
where the two masses are constant. In particular, notice
that $m_{\tilde \ph} = m / \mpl^2$, which is exactly the
expression that one encounters in supergravity for the
gravitino mass.

We thus see that the system has all the properties assumed in
section~(\ref{subgen}), so that we can apply the procedure
derived there to quantize it and to define the occupation numbers
of the fermionic eigenstates. Among the possible choices for the
transformation matrix $\Phi$ which enters into
eq.~(\ref{defphi}), we take
\begin{equation}
\Phi = \frac12  \left(\mbox{arccos} \;\; \al1 \right)
\pmatrix{\al2 / \omega & \Delta/\omega \cr
\Delta/\omega & -\al2/\omega},\label{deftil}
\end{equation}
with
\begin{equation}
\omega \equiv \sqrt{1 - \al1^2} = \sqrt{\al2^2 + \Delta^2}\,.
\end{equation}
Following eq.~(\ref{hamgen}), the hamiltonian of the
system is instead given by\footnote{Notice that the
matrix $\hat{M}_2$ that appears in the equations of
motion~(\ref{eomhat}) reads
\begin{equation}
\hat{M}_2=\left[Q,\,M\right]\,\frac{\omega}{2}+Q\,\dot{Q}\,
\frac{1-\tilde{\alpha}_1}{2}\,.
\end{equation}
At late times $\hat{M}_2\sim Q\,\dot{Q}$ does not
vanish. Indeed $\al2$ and $\Delta$ decrease at late
times in a way such that the elements of $Q$ keep on
oscillating with amplitude equal to unity. Therefore, as
discussed in the footnote before with eq.~(\ref{probh}),
the fields $\hat{X}$ are not a suitable basis for the
definition of the~hamiltonian.}
\begin{equation}
H = {\bar X} \left[ i \, \ga^i \, k_i N + {\tilde
M}_1 + \ga^0 \, {\tilde M}_2 \right] X\,,\label{hamtilde2}
\end{equation}
with
\begin{eqnarray}
{\tilde M}_1 &\equiv& M + \frac{ \omega^2}{2} \left(
Q \, M \, Q - M \right) + \frac{\partial_0 \, \al1}{2 \,
\omega} \, Q - \frac{\omega}{2} \, \al1 \, \partial_0 \,
Q\,, \nonumber\\ {\tilde M}_2 &\equiv& \frac{\omega
\, \al1}{2} \left[ M,\, Q \right] +
\frac{\omega^2}{4} \left[ Q,\, \partial_0\,Q \right].
\end{eqnarray}
We have denoted
\begin{equation}
Q = \pmatrix{\al2/\omega & \Delta/\omega \cr
\Delta/\omega & -\al2/\omega}.
\end{equation}

As we have already remarked, at late times $\al1 = - 1$,
while $\al2 = \Delta = 0$. In this regime the above
hamiltonian becomes the standard one of two decoupled
spin $1/2$~fermions
\begin{equation}
H = {\bar X} \left[ i \, \ga^i \, k_i + M \right]
X\,,\label{hamstand}
\end{equation}
with the standard gravitino mass for the field ${\tilde
\ph}$ (cf.\ eq.~(\ref{eomstand})).

We conclude this subsection discussing the explicit
diagonalization of the hamiltonian, i.e.\ of the matrices
$\overline{H}$ and $\overline{H}_0$ entering in eqs.~(\ref{eqd2})
and~(\ref{defh0}). One can now explicitly verify that the
eigenvalues of the $\overline{H}_0$ matrix occur in pairs, that
is they are of the form $\pm \omega_1, \pm  \omega_2$. One can
also verify that if $( v_1,\,v_2,\,v_3,\,v_4 )$ is an eigenvector
of $\overline{H}_0$ belonging to the eigenvalue $\omega$, \pagebreak[3] then $(
-v_3^*,-v_4^*,\,v_1^*,\,v_2^* )$ is also an eigenvector of
$\overline{H}_0$ belonging to the eigenvalue $-\omega$. We then
find
\begin{eqnarray}
R^\dagger \: \overline{H}_0 \: R = H_d\,,\qquad R =
\pmatrix{R_{1}& -R_{2}^*\cr R_{2}& R_{1}^*},\label{diah0}
\end{eqnarray}
where $H_d = \diag (\omega_1,\,\omega_2,-\omega_1, -\omega_2)$ is
the matrix that we formally introduced in eq.~(\ref{eqd1}).

The $2 \times 2$ matrices defined by eq.~(\ref{defc})
are thus given by
\begin{eqnarray}
I &=& \frac{1}{\sqrt{2}}  \left[ R_1^\dagger \, U_+ +
R_2^T \, U_- \right], \nonumber\\ J &=&
\frac{i}{\sqrt{2}}  \left[-  R_1^\dagger \, U_-^* +
R_2^T \, U_+^* \right].\label{defj}
\end{eqnarray}

We remind that the matrices $R_1$ and $R_2$ are obtained through
the diagonalization of $\overline{H}_0$ see eq.~(\ref{diah0}).
The matrices $U_+$ and $U_-$ are instead determined by their
evolution equation~(\ref{eq:3.47}). The only point left is to give
more explicitly their values at the initial time
$\overline{\eta}$. This can be done by setting $J=0$ in
eq.~(\ref{defj}), which, as we remarked, corresponds to requiring
no fermions in the initial state (see eq.~(\ref{occnum})).
Moreover, conditions~(\ref{normalgen}) have to be imposed. From
these requirements, we see that $U_+(\overline{\eta} )$ has to
fulfill
\begin{equation}
U_+^\dagger \left( \overline{\eta} \right) \left[\identity
+ R_2 \left( \overline{\eta} \right) R_1^{*-1} \left(
\overline{\eta} \right) R_1^{T-1} \left( \overline{\eta}
\right)  R_2^\dagger \left( \overline{\eta} \right) \right]
U_+ \left( \overline{\eta} \right) = 2 \, \identity\,.\label{initu}
\end{equation}
In this last expression, the matrix in square brackets is
hermitean and can be diagonalized with a unitary transformation.
More precisely, we can set it to be equal to
$V^\dagger\,\Lambda\,V$ with $\Lambda$ diagonal and real, and $V$
unitary. The initial condition for $U_+$ can thus be written
\begin{equation}
U_+\left(\overline{\eta}\right)=V^\dagger\,\sqrt{2\,
\Lambda^{-1}}\,.
\end{equation}
Finally, $U_-(\overline{\eta})$ is obtained by setting $J (
\overline{\eta} ) = 0$ in eq.~(\ref{defj}).

\subsection{Analytical results with unbroken supersymmetry in
the vacuum} \label{sub4}

The case ${\hat \mu}^2 = 0$ is particularly interesting
since some results can be worked out analytically, and
since it provides some hints between the final gravitino
abundance and the size of supersymmetry breaking. For
${\hat \mu}^2 = 0$ supersymmetry is unbroken in the
minimum of the theory, at $\phi = s =
0$.\footnote{Indeed for ${\hat \mu}^2$ strictly zero the
potential for the Polonyi field becomes flat for $\phi =
0$, and $s \equiv 0$ for the whole evolution.} Because
of this, in the vacuum of the theory the gravitino has
only the transverse component.

The computation of the formulae of section~\ref{sub3} is in this
case particularly simplified. The quantity $\Delta$ vanishes
identically, so that the two fields $\Upsilon$ (which is always
the Polonyi fermion) and $\ph$ (which is always the inflatino)
are decoupled. Going back to the formalism of
section~(\ref{subgen}), we find that we have to perform only the
first redefinition of the fermions, $X
\equiv \exp ( - \ga^0 \, \Phi ) \, {\hat X}$,
where now $\Phi = \diag (
\varphi,- \varphi )$, with
\begin{equation}
\cos \left( 2 \, \varphi \right) = \al1\,,\qquad
\sin \left( 2 \, \varphi \right) = \al2\,.
\end{equation}
The two redefined fields have the ``standard'' equations
of motion and hamiltonian
\begin{eqnarray}
\left( \ga^0 \, \partial_0 + i \, \ga^i \, k_i +
m_{\hat \ph} \right) {\hat \ph} &=& 0\,, \nonumber\\
\left( \ga^0 \, \partial_0 + i \, \ga^i \, k_i +
m_{\hat \Upsilon} \right) {\hat \Upsilon} &=& 0\,,
\nonumber \\ H &=& \int d^3 \mathbf{k} \left[ {\bar {\hat
\ph}} \left( i \, \ga^i \, k_i + m_{\hat \ph} \right)
{\hat \ph} + {\bar {\hat \Upsilon}} \left( i \, \ga^i \,k_i
+ m_{\hat \Upsilon} \right) {\hat \Upsilon} \right], \nonumber\\
m_{\hat \ph} &=& m_\ph + \partial_0\varphi\,,\qquad m_{\hat
\Upsilon} = m_\Upsilon -\partial_0 \varphi\,,
\end{eqnarray}
with $m_\ph$ and $m_\Upsilon$ given in
eq.~(\ref{masstilde}).

In practice, ``removing'' the time dependent matrix
which multiplies the momentum in the original equations
for $\ph$ and $\Upsilon$ gives an additional
contribution to the mass of the fields ${\hat \ph}$ and
${\hat \Upsilon}$, as noticed in~\cite{kklv1,grt1}. When
$\Delta = 0$, an explicit computation of $\partial_0
\phi$ with the present formalism is provided
in~\cite{kklv2}
\begin{equation}
\partial_0 \varphi = - a \, \frac{\dot{\al2}}{2 \, \al1} = a \,
\left( m_{11} + m \right) + 3 \, a \, \left( H \,
\frac{\dot{\phi}}{\sqrt{2}} - m_1 \, m \right)
\frac{m_1}{\frac{1}{2} \, \dot{\phi}^2 + m_1^2}\,,
\end{equation}
where the various quantities have been introduced in
section~\ref{sub1}. Using the equation of motion for the inflaton
field $\phi$, one can show that it is precisely $\partial_0
\varphi \equiv m_\Upsilon$, so that the field ${\hat \Upsilon}$
is effectively massless. The mass for ${\hat \ph}$ is instead of
the order the inflaton mass. More precisely, it has a variation
of the order $m_\phi$ within the first oscillation of the
inflaton (i.e.\ in the time $m_\phi^{-1}$) and then it stabilizes
at $m_\phi$~\cite{kklv1,grt1}. Since the fields are decoupled,
the formulae for the occupation numbers~(\ref{eq1ferm}) are quite
simple. They show that the Polonyi fermion is not produced, while
the production of the inflatino field has a cut-off at $k
\sim m_\phi$, and decreases as $k^{-4}$ at
big~momenta.\footnote{This can be explicitly seen by
integrating eq.~(\ref{eq1ferm}) for $\dot{B}$ in the
limit of large $k$ and with $A \simeq 1$.}

The main point of this subsection is that the Polonyi
fermion is \emph{not} produced at preheating for ${\hat
\mu}^2$ strictly zero. When ${\hat \mu}^2 \neq 0$ the
Polonyi fermion provides the longitudinal component for
the gravitino, so its abundance turns out crucial to
understand whether gravitinos are or are not
overproduced. If one believes that the limit ${\hat
\mu}^2 \rightarrow 0$ is continuous, the present
analysis suggests indeed that the production of
gravitinos should become smaller as ${\hat \mu}^2$
decreases. Although we do not have a rigorous proof of
this continuous behavior,\footnote{The problem is that
the dynamics of the Polonyi field is governed by the
timescale ${\hat \mu}^{-2} \, m_\phi^{-1}$, which
becomes infinite in the limit ${\hat \mu}^2 \rightarrow
0$~\cite{kklv2}.} the numerical results that we show in
the next subsection strongly support this assumption.

\subsection{Numerical results with broken supersymmetry in the
vacuum} \label{sub5}

We now analyze the situation ${\hat \mu}^2 \neq 0$. As
we have said, in this case the quantity $\Delta$ is
generally also non vanishing in the most interesting
part of the evolution. As a consequence, the dynamics of
the fermionic fields ${\tilde \ph}$ and ${\tilde
\Upsilon}$ is coupled, i.e.\ we have mixed terms in
their equations of motion~(\ref{eomtilde}) and in their
hamiltonian~(\ref{hamtilde2}). For the following
discussion it is useful to explicitly write ${\tilde
\Upsilon}$ in terms of the chiral fields $\chi_1$ and
$\chi_2$. Combining the
definitions~(\ref{manydef}),~(\ref{defdel}),
and~(\ref{deftil}) we have, for minimal K\"ahler
potential and real scalar fields,
\begin{equation}
{\tilde \Upsilon} = \frac{a^{3/2} [ m_1^2 + m_2^2 +
\frac{1}{2} \, \dot{\phi}_1^2 + \frac{1}{2} \,
\dot{\phi}_2^2 ]^{1/2}}{m_1 \, \dot{\phi}_2^2 -
m_2 \, \dot{\phi}_1^2} \left( \dot{\phi}_1 \, \chi_1
+\dot{\phi}_2 \, \chi_2 \right),\label{yexp}
\end{equation}
where we remind that the two scalars $\phi_1$ and
$\phi_2$ are the inflaton and the Polonyi field, while
$\chi_1$ and $\chi_2$ the corresponding fermions.
Moreover, the definition~(\ref{gold}) of the goldstino
now reads
\begin{equation}
\upsilon_L = \left( m_i - \frac{1}{\sqrt{2}} \, \ga^0 \, a \,
\dot{\phi}_i \right) \chi_i\,.
\end{equation}

Let us first consider the initial and final stages of
the evolution, where only one of the two scalar fields
significantly contribute to the supersymmetry breaking
and the quantity $\Delta$ is practically vanishing.
During inflation, one has $m_2,\, \dot{\phi}_1,\,
\dot{\phi}_2 \simeq 0$, and the supersymmetry breaking
is provided almost completely by $m_1$. Moreover, the
goldstino is practically the field $\chi_1$. We remind
that we are working in the unitary gauge, so that
$\chi_1 \simeq \upsilon = 0 $. Equation~(\ref{yexp})
thus rewrites
\begin{equation}\label{yin}
{\tilde \Upsilon} \simeq \frac{a^{3/2} \, \vert m_1
\vert}{m_1 \, \dot{\phi}_2} \, \chi_1 \, \dot{\phi}_2 =
a^{3/2} \, \chi_1\,,\qquad {\hat t} \ll {\hat \mu}^{-2}\,.
\end{equation}
Notice the factor $a^{3/2} $ appearing in the last
expression, which is a consequence of the fact that the
field ${\tilde \Upsilon}$ is canonically normalized in
comoving units (cf.\ the discussion after the
lagrangian~(\ref{lagnoi})).\footnote{One may also worry
about dividing by $\dot{\phi}_2 \simeq 0$ in
eq.~(\ref{yin}). However, this is due to the fact that
the quantity $\Upsilon$ is ill --- defined in the static
$\dot{\phi}_i \rightarrow 0$ limit, while ${\tilde
\Upsilon}$ is not.} In the late stages of the evolution,
supersymmetry is instead broken by the Polonyi field,
and the only non-vanishing contribution is provided by
$m_2$. With the same arguments used to get
eq.~(\ref{yin}), one can show~that\footnote{One can also
show that, in the late stages of the evolution, $\vert
\dot{\phi}_1 \vert \simeq \vert \dot{\phi}_2 \vert$.}
\begin{equation}
\upsilon \propto \chi_2 = 0\,,\qquad {\tilde \Upsilon} =
- a^{3/2} \,\chi_1\,,\qquad {\hat t} \gg {\hat \mu}^{-2}\,.
\end{equation}

In order to compute the evolution of the occupation
number of the fermions we adopt the procedure described
in section~\ref{subgen}: during the evolution of the
system, the states are mixed in such a way that the
hamiltonian is kept in a diagonal form. The two
eigenstates obtained through this diagonalization
coincide with the fields $\ph$ and $\Upsilon$ only for
$\Delta = 0$. In particular, this is true at very early
and late times. The safest way to make the proper
identifications in these regimes is to consider the
evolution of the mass eigenvalues, which always behave
like in the example shown in figure~\ref{fig3}. The two
masses present (for ${\hat \mu}^{-2} \ll 1$) a strong
hierarchy. We denote with $\psi_1$ the eigenstate with
bigger mass, and with $\psi_2$ the other one. The mass
of $\psi_1$ converges to the inflatino mass ($\simeq
1.31~m_\phi$) at late times, and it is always of the
order $m_\phi$. On the contrary, the mass of $\psi_2$
converges to the gravitino mass ($\simeq
1.31~\hat{\mu}^2\,m_\phi$) in the vacuum. As it will be
clear below, we can ``qualitatively'' identify $\psi_1$
with the inflatino and $\psi_2$ with the Polonyi fermion
for the whole evolution. Although rigorous only at late
times, this identification can be useful for a
qualitative understanding of the system.

\FIGURE[t]{\epsfig{file=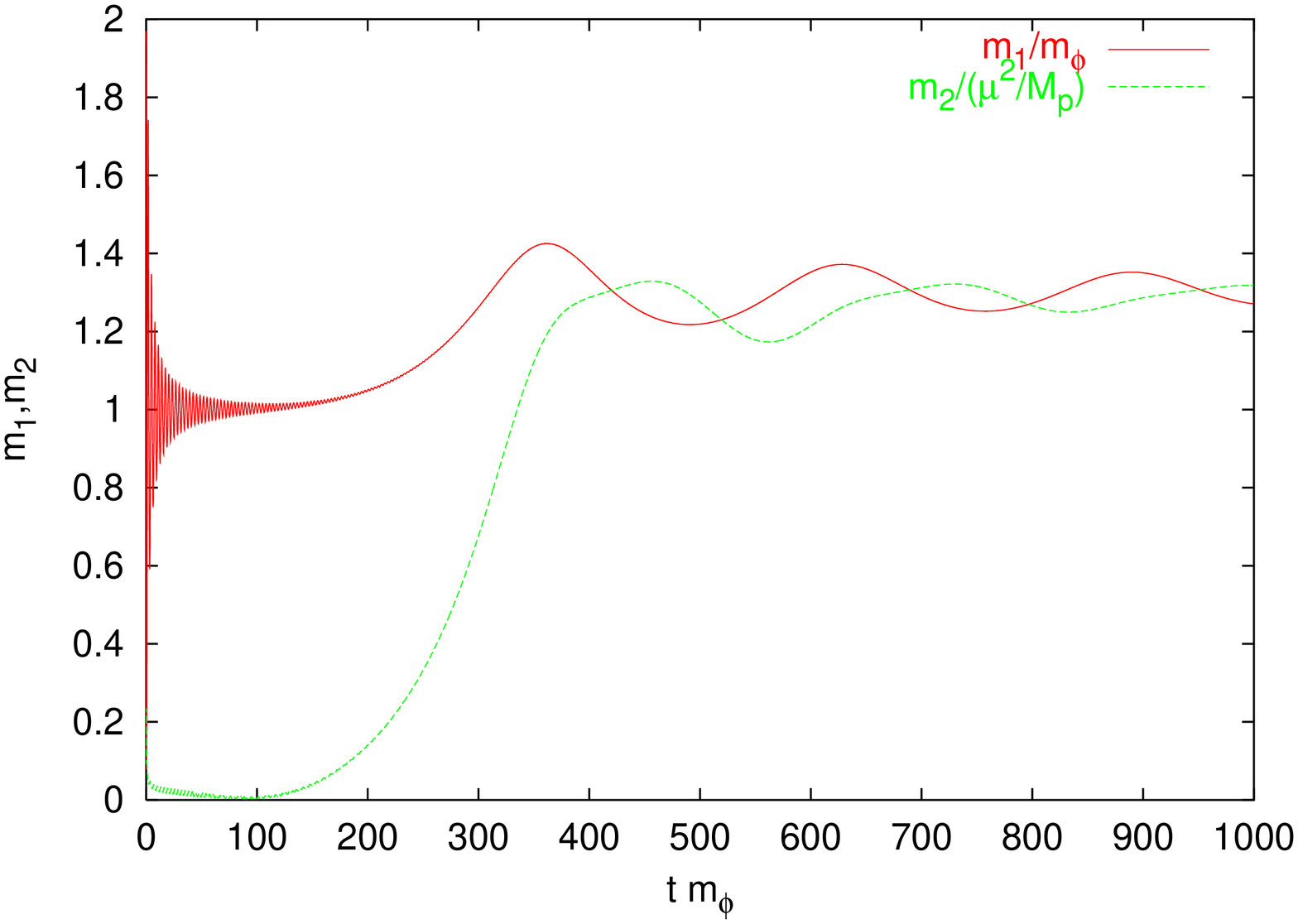,angle=-90,width=0.8 \textwidth}
\caption{Evolution of the masses of the two fermionic eigenstates.
As in
figure~\ref{fig1}, ${\hat \mu}^2 = 10^{-2}$. Notice
the different normalizations for the two masses.}
\label{fig3}}

What is most important to us is the relation between the
eigenstates $( \psi_1,\, \psi_2 )$ and the gravitino $\ph$ and
the matter field $\Upsilon$. As we have said, the last fields
coincide with the physical eigenstates only at the very beginning
and at the end of the evolution. More precisely, from the
behavior of the two masses we have
\begin{equation}
\ph \equiv \psi_2 \qquad\mbox{and} \qquad \Upsilon
\equiv \psi_1\label{idem1}
\end{equation}
at late times. On the contrary, it must be
\begin{equation}
\ph \equiv \psi_1 \qquad \mbox{and} \qquad \Upsilon
\equiv \psi_2\label{idem2}
\end{equation}
at early times, since the longitudinal gravitino
component $\ph$ is provided by the goldstino and
supersymmetry is initially broken only by the inflaton
field.

At intermediate times, the hamiltonian cannot be
diagonalized with a simple rotation in ``flavor'' space,
and $\ph$ cannot be just a simple (i.e.\ with only
numbers as coefficients) linear combination of $\psi_1$
and $\psi_2$. However, we can gain an intuitive
description of the system through the identifications
\begin{eqnarray}
\ph &\sim& \sqrt{r_\phi} \, \psi_1 + \sqrt{r_s} \, \psi_2\,,
\nonumber\\
\Upsilon &\sim& - \sqrt{r_s} \, \psi_1 + \sqrt{r_\phi} \,
\psi_2\,.\label{upth12}
\end{eqnarray}
The coefficients $r_\phi$ and $r_s$ give a ``measure''
of the relative contribution to supersymmetry breaking
provided by the two scalar fields (see
eq.~(\ref{defrs})). These relations can thus be
justified as a ``generalization'' of the equivalence
theorem, in a way also suggested in~\cite{grt2,kklv2}.
We remark that they are rigorous at early and late times
(when they coincide with the
identifications~(\ref{idem1}) and~(\ref{idem2})). At
intermediate times they interpolate between these two
regimes and can be thus used as a qualitative
description of the system.

From eqs.~(\ref{upth12}) we deduce the following
estimates for the occupation numbers
\begin{eqnarray}
N_\ph &=& r_\phi \, N_1 + r_s \, N_2\,, \nonumber\\
N_\Upsilon &=& r_s \, N_1 + r_\phi \, N_2\,.
\end{eqnarray}
The evolution of these quantities is shown in
figure~\ref{fig4} for modes of comoving momentum
$k=m_\phi$ and for ${\hat \mu}^2 = 10^{-2}$. Notice
that (by construction) $N_\ph \equiv N_1$ at early
times, while $N_\ph\equiv N_2$ at late ones. In these
regimes these identifications are rigorous.

\FIGURE[t]{\epsfig{file=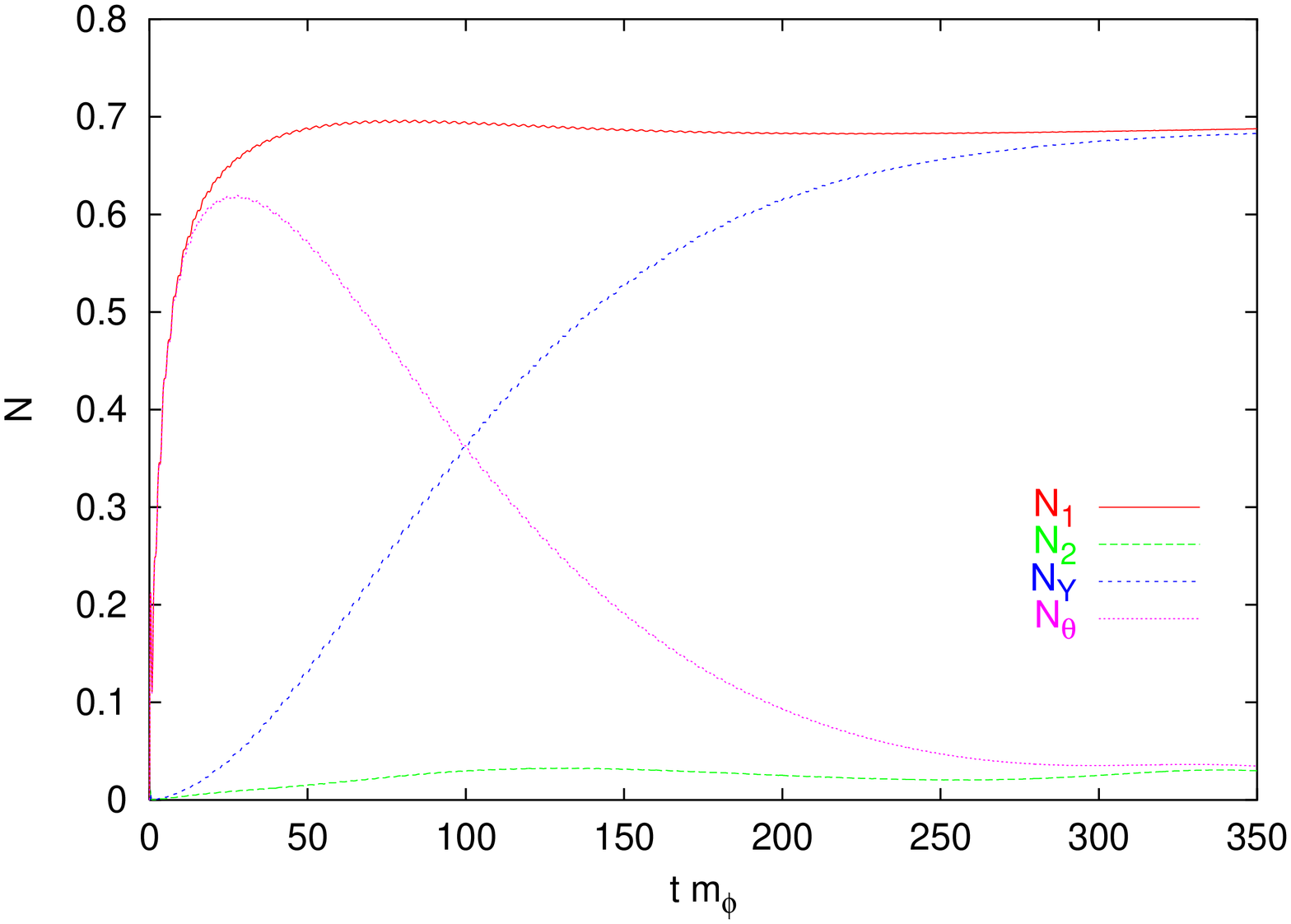,angle=-90,width=0.8 \textwidth}
\caption{Evolution of $N_\theta$ and $N_\Upsilon$ for
${\hat \mu}^2 = 10^{-2}$ and $k = m_\phi$. See the
text for details.}\label{fig4}}

In figures~\ref{figev1} and~\ref{figev2} we plot instead
the spectra of the states $\psi_1$ and $\psi_2$ in the
case ${\hat \mu}^2 = 10^{-2}$ and at the times $t = 10
\, m_\phi^{-1}$ (that is, after a couple of
oscillations of the inflaton), $t = \hat{\mu}^{-2} \,
m_\phi^{-1}$, and $t = 10 \, \hat{\mu}^{-2} \,
m_\phi^{-1}$.

\FIGURE[t]{\epsfig{file=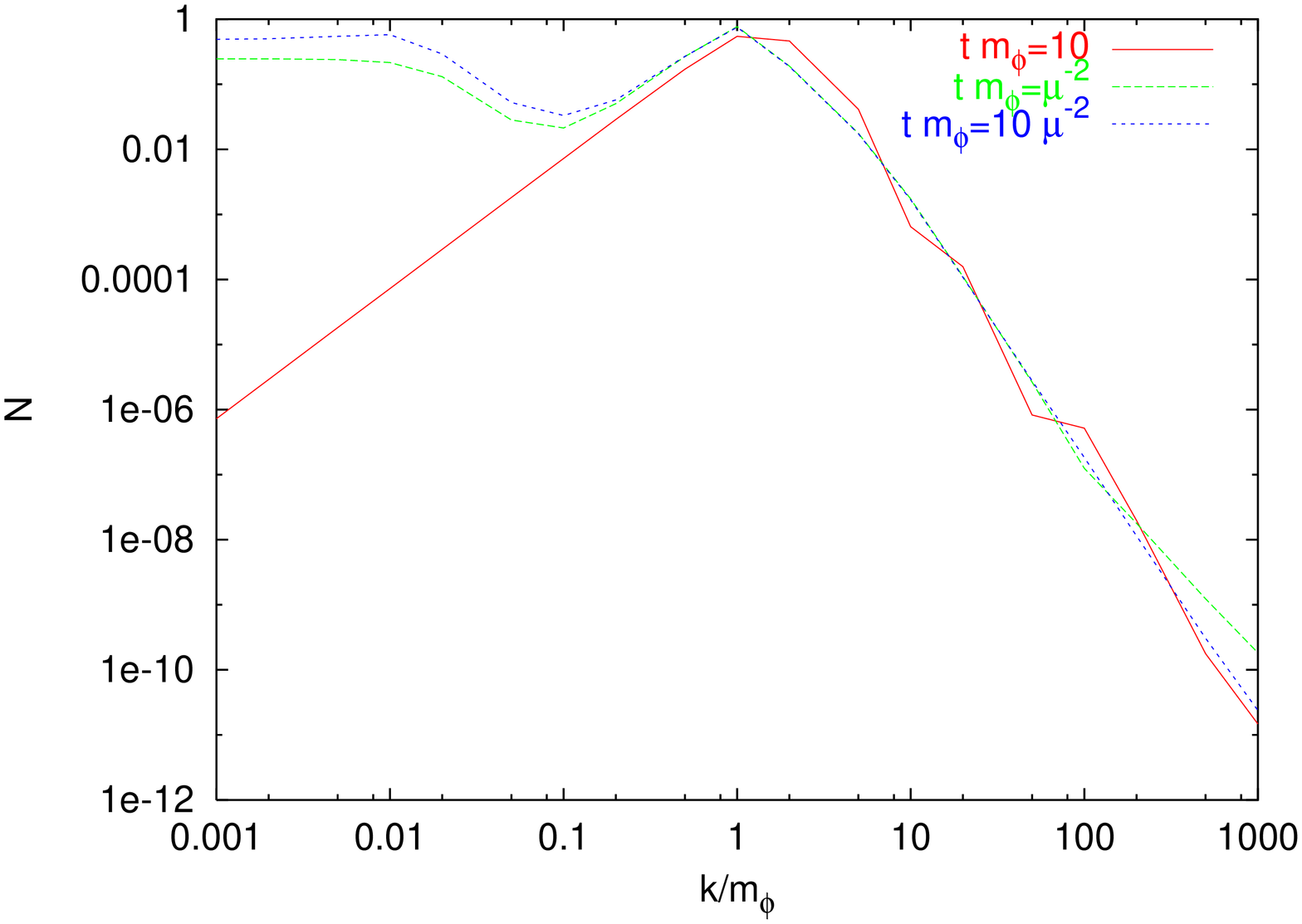,angle=-90,width=0.8 \textwidth}
\caption{Spectrum of the state $\psi_1$  at different times for
$\hat{\mu}^2=10^{-6}$.}\label{figev1}}

\FIGURE[t]{\epsfig{file=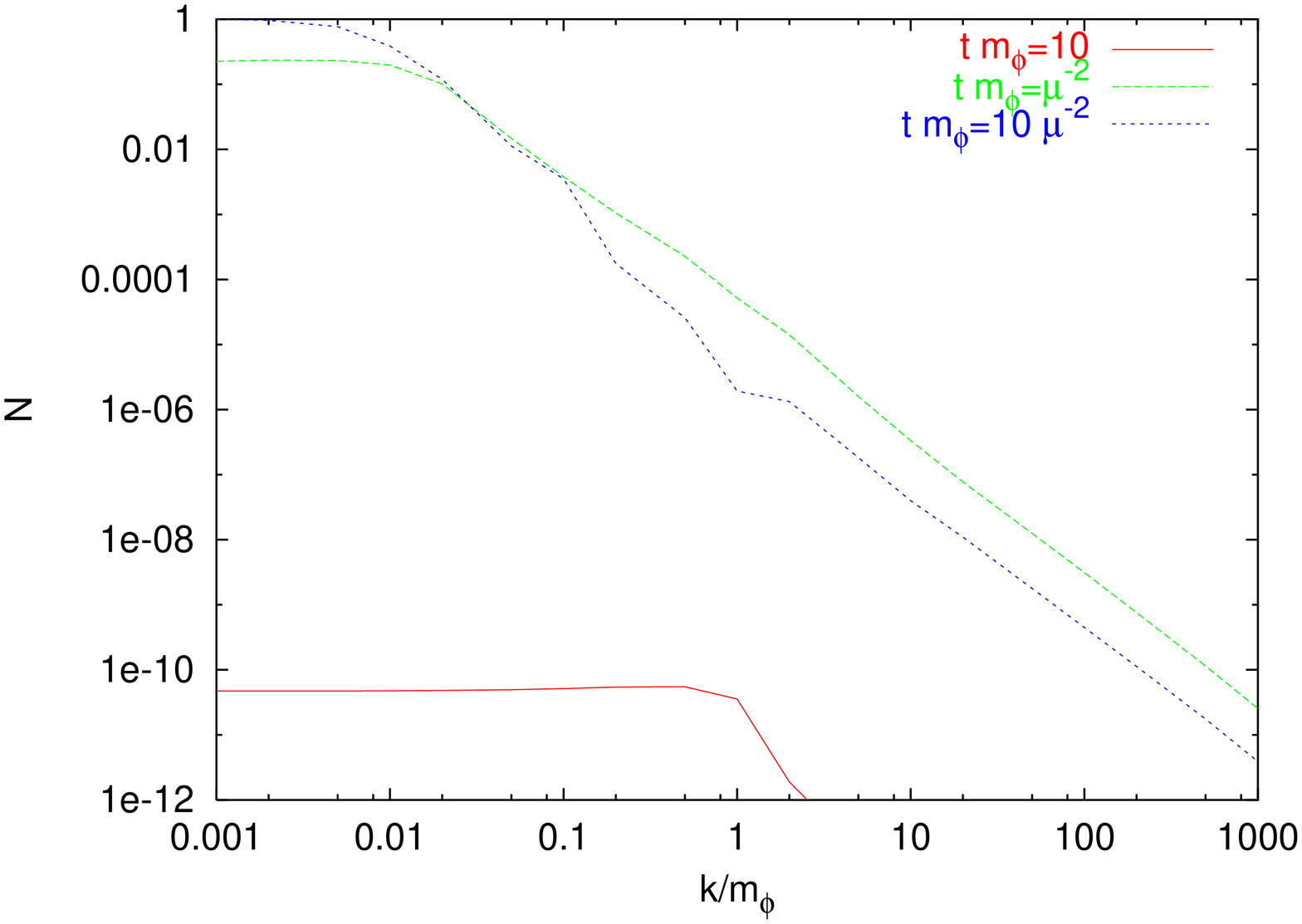,angle=-90, width=0.8\textwidth}
\caption{Spectrum of the state $\psi_2$  at different times for
$\hat{\mu}^2=10^{-6}$.}\label{figev2}}

It is apparent that most quanta of the state $\psi_1$ are
produced at the very first oscillations of the inflaton field,
while quanta of $\psi_2$ are mainly produced at the times when
the Polonyi scalar starts oscillating. This supports the
qualitative identification of $\psi_1$ with the inflatino and of
$\psi_2$ with the Polonyi fermion. It is worth noticing that, for
comoving momenta smaller than $m_\phi$, the increase of $N_2$ is
not related to a ``conversion'' of quanta of $\psi_1$ to
$\psi_2$. Indeed, the increase in $N_2 ( k )$ is not accompanied
by a decrease in $N_1 ( k )$ for $k \lta m_\phi$.

We can now show our most important result, that is the
spectra of $\Upsilon$ and $\ph$ at the end of the
process. We present them in figures~\ref{fig5}
and~\ref{fig6}, respectively. They are computed at the
time\footnote{In the cases ${\hat \mu} = 10^{-2} -
10^{-4}$ we have continued the evolution further,
until the spectra stop evolving. We have found that the
spectra shown in figure~\ref{fig5} coincide with the
final ones, while $N_\theta$ very slightly decreases for
$t > 10 \, {\hat\mu}^{-2} m_\phi^{-1}$. Thus, we
believe the results shown in figure~\ref{fig6} to
provide an accurate upper bound on the final gravitino
abundance.} $t=10 \, {\hat\mu}^{-2} m_\phi^{-1}$.
The time required for the numerical computation
increases linearly with ${\hat \mu}^2$, and the
realistic case ${\hat \mu}^2 = 10^{-11}$ is far from
our available resources. We thus kept $\hat\mu^2$ as a
free parameter and we performed the explicit numerical
computation only up to $\hat{\mu}^2=10^{-6}$. In
particular, in figures~\ref{fig5} and~\ref{fig6} the
spectra of the fermions produced at preheating are shown
for $\hat{\mu}^2 = 10^{-2},\, 10^{-3},\,
10^{-4},\,10^{-5},\,10^{-6}$. The case ${\hat \mu}^2 =
10^{-11}$ can be clearly extrapolated from the ones
shown in these figures.

\FIGURE[t]{\epsfig{file=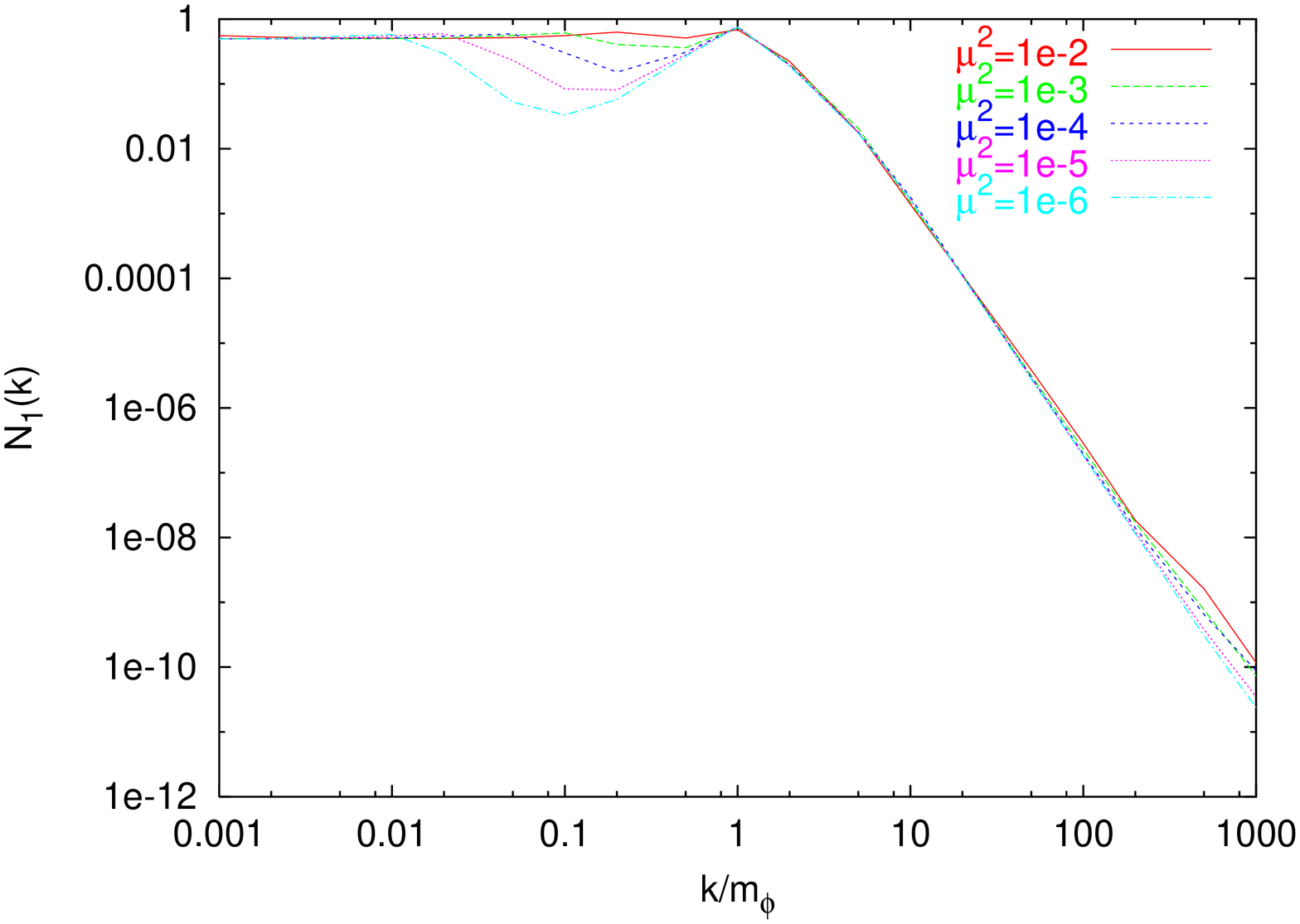, angle=-90, width=0.84\textwidth}
\caption{Spectrum of inflatinos at late times.} \label{fig5}}

\FIGURE[t]{\epsfig{file=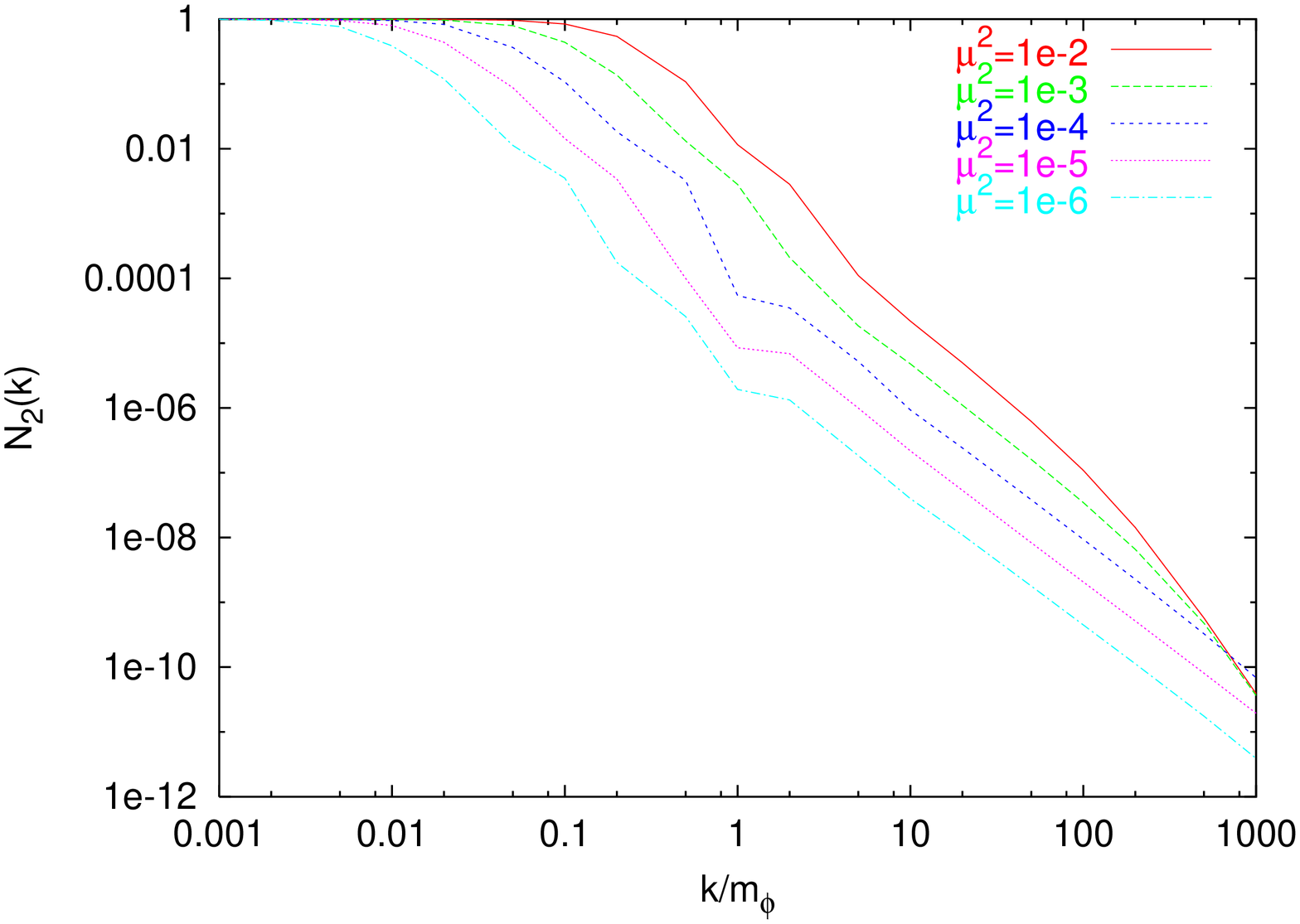, angle=-90, width=0.8\textwidth}
\caption{Spectrum of gravitinos at late times.}
\label{fig6}}

In figure~\ref{fig5}, the spectra for the state $\psi_1$
are shown. This state corresponds to the matter fermion
$\Upsilon$ in the true vacuum. It is apparent that the
main features of the spectrum are independent of the
value of $\hat{\mu}^2$. The reason for this is that
$\psi_1$ is associated to the inflatino, that is
produced by the coherent oscillations of the inflaton.
As we discussed above, the dynamics responsible for the
production of this state is independent on the value of
$\hat{\mu}^2$. Therefore the only relevant scale for
$N_1$ is $m_\phi$, and indeed the spectrum of $\psi_1$
exhibits a cut-off at comoving momentum $k\sim m_\phi$.\pagebreak[3]

The spectra shown in figure~\ref{fig6} are related to the
abundance of gravitinos after the fields have stabilized in their
minima. We see that in this case the occupation number decreases
as $\hat{\mu}^2$ becomes smaller. Indeed, the occupation number
$N_2(k)$ is of order unity for comoving momenta $k$ smaller than
some cut-off $k_*$. From figure~\ref{fig6} we can deduce the
dependence $k_*\propto ( \hat{\mu}^2 )^{1/3}$ of this cut-off on
the parameter ${\hat \mu}^2$. This behavior suggests that
gravitinos are not produced in the limit ${\hat \mu}^2
\rightarrow 0$, confirming what we have argued in the
section~\ref{sub4}.

We can conclude that in the model we are considering
both inflatinos and gravitinos are produced
nonthermally. However, the mechanism responsible for the
production of gravitinos is much less efficient that the
one acting on inflatinos. Indeed, the latter is related
to the dynamics of the inflaton, while the former is
related to the dynamics of the Polonyi field. As a
consequence, and as it is clearly confirmed by the
comparison of figures~\ref{fig5} and~\ref{fig6}, the
number of non-thermal gravitinos is much smaller than the
number of inflatinos.

\section{Conclusions}

Particles with gravitational decay constitute one of the
most serious danger for cosmology, since they can spoil
the successful predictions of primordial
nucleosynthesis. This is particularly true for
gravitinos in models where supersymmetry is
gravitationally broken. To avoid this danger, strong
upper limits must be imposed on their number. While
gravitino thermal production has been very well studied
and understood over the last twenty years, only very
recently non-perturbative creation after inflation has
been considered. Non-perturbative production appears more
involved than the thermal one, and indeed explicit
calculations were so far performed only in the one
superfield case with supersymmetry unbroken in the
vacuum. The results of these calculations showed that
the coherent oscillations of the inflaton field easily
cause an overproduction of the longitudinal gravitino
component. However, this component is absent in the
vacuum of the theory if supersymmetry is unbroken. One
is thus led to wonder if these quanta are really
dangerous gravitinos or should be better understood as
harmless inflatinos (in case of only one chiral
multiplet, the scalar is necessary the inflaton). In
order to discriminate between inflatino and gravitino
production it is necessary to consider more realistic
schemes. The simplest natural possibility is to consider
two separate sectors, one of which drives inflation,
while the second is responsible for supersymmetry
breaking today. In the present analysis we have studied
the situation in which the two sectors communicate only
gravitationally, in the hope that suppressing their
interaction may lead to a small gravitino production.

The presence of more than one chiral superfield
increases the difficulty of the problem. One has to deal
with a (quite involved) coupled system, constituted by
the gravitino longitudinal component and some
combinations of the matter fields, in the external
background made by the scalar fields of the theory. The
most difficult part of this analysis is to provide a
formalism in which the coupled system is quantized, with
a clear definition of the occupation numbers for the
physical eigenstates. This is a very interesting problem
by itself, which can have several other applications
besides the gravitino production. We faced this problem
in the first part of the work. We showed that the
standard procedure for quantizing and providing the
occupation number for one field can be generalized to
systems of multi coupled fields, both in the bosonic and
in the fermionic case. Although far from trivial, this
generalization can be presented in a remarkably simple
form. The application of this formalism to the gravitino
production is performed in the last section of the
paper. We have shown that, in the specific model
considered here, the number of produced gravitinos is
very sensitive to the size of the final supersymmetry
breaking, and that it actually vanishes in the limit in
which supersymmetry is preserved. Due to the small scale
of the expected supersymmetry breakdown, we conclude
that gravitino non-thermal production is very suppressed
in this model.

\paragraph{Note added.}

Contemporarily to the present manuscript, the
work~\cite{grt3} appeared on the database. This work
does not consider gravitinos, but studies the non-thermal
production of moduli fields coupled to the inflaton
sector. Moduli fields can also be very dangerous for
cosmology because, analogously to gravitinos, they are
expected to decay gravitationally. The work~\cite{grt3}
shows that moduli production can be significant if ---
after inflation --- they are strongly coupled to the
inflaton. These considerations seem to agree with our
suggestion~\cite{noi} that strong couplings between
potentially dangerous relics (in our case gravitinos)
and the inflaton sector should be avoided in any viable
cosmological model.

\acknowledgments
We are pleased to thank R.~Kallosh, L.~Kofman, A.~Linde,
D.H.~Lyth, S.~Petcov, A.~Riotto, and A.~Van Proeyen for
interesting conversations. This work is supported by the
European Commission RTN programmes HPRN-CT-2000-00148
and 00152.

\catcode`\%=12
\newcommand\asj[3]
{\href{http://www-spires.slac.stanford.edu/spires/find/hep/www?j=ASJOA%2C#1%2C#3
}
                {{\it Astrophys.\ J.\ }{\bf #1} (#2) #3}}
\catcode`\%=14

\renewcommand\baselinestretch{1.08}

\end{document}